\documentclass{ieeeaccess}

\usepackage[T1]{fontenc}
\usepackage[utf8]{inputenc} 
\usepackage{cite}
\usepackage{amsmath,amssymb,amsfonts}
\usepackage{graphicx}
\usepackage{textcomp}
\usepackage{xcolor}
\usepackage{booktabs}
\usepackage{tabularx}
\usepackage{array}
\usepackage{xspace}
\usepackage{url}

\makeatletter
\DeclareRobustCommand\onedot{\futurelet\@let@token\@onedot}
\def\@onedot{\ifx\@let@token.\else.\null\fi\xspace}

\def\eg{\emph{e.g}\onedot}

\def\etal{\emph{et al}\onedot}
\makeatother

\makeatletter
\providecommand\xfigwd{0pt}
\makeatother

\begin{document}
\history{Date of publication xxxx 00, 0000, date of current version xxxx 00, 0000.}
\doi{10.1109/ACCESS.2024.DOI}

\title{A Configuration-First Framework for Reproducible, Low-Code Machine
Learning: a Localization Use Case}

\author{TIM STRNAD,
BLA\v{Z} BERTALANI\v{C},
AND CAROLINA FORTUNA}

\address{Jo\v{z}ef Stefan Institute, Jamova cesta 39, 1000 Ljubljana,
Slovenia (e-mail: tim.strnad@ijs.si; blaz.bertalanic@ijs.si;
carolina.fortuna@ijs.si)}

\tfootnote{This work was supported by the Slovenian Research and Innovation
Agency (ARIS) under Programme P2-0016, and by the European Union through the
NANCY Project under Grant Agreement No. 101096456 and the ACCEPTREMOTE Project
under Grant Agreement No. 101269278.}

\markboth
{Strnad \headeretal: A Configuration-First Framework for Reproducible, Low-Code Machine Learning}
{Strnad \headeretal: A Configuration-First Framework for Reproducible, Low-Code Machine Learning}

\corresp{Corresponding author: Carolina Fortuna (e-mail: carolina.fortuna@ijs.si).}

\begin{abstract}
    As machine learning underpins more critical applications, the value of a reported result depends on whether it can be compared and repeated. In practice, this remains difficult: research groups often assemble their own tools for configuration, execution, versioning, and evaluation, while also repeating the domain-specific work such as dataset preparation and baseline implementation. We present a configuration-first design for application-specific ML experimentation frameworks that addresses these sources of repeated effort. An experiment is declared in human-readable configuration files; a workflow orchestrator executes each stage as an isolated process that communicates only through explicit inputs and outputs; and code, data, configurations, environment specifications, and generated artifacts are versioned together, so that a recorded run can be inspected and repeated. We instantiate the design as LOCALIZE for radio-localization research, which supplies preconfigured datasets, processing stages, model-development procedures, and experiment templates while leaving the underlying pipeline open to modification. A qualitative comparison against five experimentation platforms, together with controlled quantitative studies against matched Jupyter notebook and Kedro implementations, shows that for the localization workflows studied, changes covered by LOCALIZE's supplied components require fewer codebase edits, while total wall-clock time and peak memory usage remain comparable. In a controlled scaling experiment at \(1\times\), \(5\times\), and \(10\times\) the base dataset volume, total CPU and wall time grew sublinearly over the tested sizes.
\end{abstract}

\begin{keywords}
    Experiment management, localization, low-code, machine learning,
    reproducibility, workflow orchestration.
\end{keywords}

\titlepgskip=-15pt

\maketitle

\section{Introduction}
\label{sec:intro}
\PARstart{B}{ackground and motivation.} Machine learning (ML), both classical and deep, now underpins applications in domains from logistics to healthcare, and with that reliance comes a reproducibility problem in the research software that produces ML results~\cite{Pineau2021, Heil2021}. Model development itself remains largely ad hoc: codebases are fragile, and small differences in global state or execution environment change the reported outcome~\cite{bouthillier_accounting_2021}. \textit{Radio-based localization} is one field in which this matters, because it underpins navigation, logistics, social, entertainment, and healthcare services~\cite{huang2018location}, and because it is moving from analytical models to learned ones~\cite{trevlakis_localization_2023}. That shift carries the reproducibility problem into localization research, so the tools and pipelines used there must follow.

\textit{Gaps in current practice.} Data-driven pipelines have produced results across the sciences, including the automatic discovery of physical laws~\cite{makke2024interpretable}, but they also strain the practices used to produce and check those results~\cite{Heil2021} and raise development problems that are not anticipated at design time~\cite{dolata2023making}. Without standardized and extensible workflows, effort is duplicated, results are fragile, and domain experts who are not also software engineers face a high barrier to entry~\cite{wilson2017good}. The two dominant responses trade against each other: GUI and low-code platforms cut development effort but constrain advanced experimentation and fine-grained control~\cite{frank_weka_workbench_2016,orange_pdf,NETIoT,DriotData,zhuang_easyfl_2022}, whereas code-centric workflows impose no such limits but, absent dedicated tooling, leave reproducibility, environment control, and consistent evaluation to the researcher~\cite{Pineau2021,Heil2021,henderson_deep_2018,bouthillier_accounting_2021,gundersen2021do}. Neither response removes the setup cost. In much of the localization literature~\cite{trevlakis_localization_2023}, as elsewhere, the contribution is a new model or training method evaluated on public datasets, so preparing those datasets and reimplementing baselines is overhead rather than the object of study. That overhead is not merely tedious. Preprocessing choices, hyperparameters, library versions, and random seeds can each move the conclusion and defeat fair comparison~\cite{Pineau2021, Heil2021}, and exact repeatability is more fragile still: hidden global state, non-deterministic operations, data-order effects, and hardware differences alter outputs even when seeds and settings are fixed~\cite{bouthillier_accounting_2021, henderson_deep_2018, pham_problems_nodate}.

\textit{Research questions.} These gaps motivate four research questions.

\begin{itemize}
    \item \textbf{RQ1} How can a configuration-first, application-specific ML experimentation platform be designed and instantiated?
    \item \textbf{RQ2} How do the codebase changes required for incremental experiment modifications compare with representative alternatives?
    \item \textbf{RQ3} What are the computational resource trade-offs compared to representative alternatives?
    \item \textbf{RQ4} How does the resource use of the proposed platform change with dataset volume?
\end{itemize}

\textit{Proposed approach.} We treat this as a design question: what must an ML experimentation environment provide so that (i) results are repeatable by default, (ii) routine work requires configuration rather than code, and (iii) stages stay independent enough for advanced experimentation? Our answer is a configuration-first design (RQ1) that fixes the conditions surrounding each stage while leaving the behavior of the stage code itself to the user, an instance of engineering within boundaries. We instantiate and evaluate this design for a radio-localization use case as LOCALIZE\footnote{\url{https://github.com/sensorlab/localize}}, which supplies preconfigured datasets, processing stages, model-development procedures, and a shared evaluation layer for the domain.

\textit{Contributions.} This paper makes the following contributions:
\begin{itemize}
    \item
          A configuration-first design for application-specific ML experimentation frameworks (RQ1). Experimental intent is declared in configuration and kept separate from execution, which is isolated per stage; code, data, configurations, environment specifications, and artifacts are versioned together, so any recorded run can be inspected and repeated within the same controlled environment.
    \item
          LOCALIZE, an instantiation of the design for radio localization (RQ1), packages the datasets and experiment procedures the field uses repeatedly, and leaves users free to override the configuration, an individual component, or a whole stage.
    \item
          An evaluation combining a qualitative comparison against five ML experimentation platforms (RQ1) with controlled quantitative comparisons against Jupyter notebook and Kedro implementations. For the localization workflows studied, changes already covered by LOCALIZE's supplied components need fewer codebase edits (RQ2), while total wall-clock time and peak memory remain comparable across repeated runs (RQ3). We further characterize how LOCALIZE's resource use grows as dataset volume rises from \(1\times\) to \(10\times\) (RQ4).
\end{itemize}

Section~\ref{sec:related} reviews related work, Section~\ref{sec:problem} states the problem and derives the design requirements, Section~\ref{sec:framework} presents the framework, and Section~\ref{sec:implementation} its reference implementation.  Sections~\ref{sec:methodology} and~\ref{sec:evaluation} give the evaluation methodology and results, and Section~\ref{sec:conclusion} the conclusion.

\section{Related Work}
\label{sec:related}
We review four bodies of work: evidence on why ML results are hard to reproduce, general-purpose experimentation platforms, experiment-management and operational infrastructure, and application-specific research software. The first establishes what a solution must control; the remaining three supply mechanisms that the proposed design combines.

\subsection{The Challenge of Reproducibility in Machine Learning}
ML results are sensitive to choices that appear inconsequential, so reproducibility has two distinct aspects: how much results vary across runs, and whether a run repeats exactly under fixed conditions~\cite{semmelrock_reproducibility_2025,bouckaert_estimating_2004}. Henderson \etal \cite{henderson_deep_2018} show that non-determinism in benchmark environments, together with variance caused by experimental and implementation choices, can materially affect reported results, and call for significance testing and standardized reporting. Bouthillier \etal \cite{bouthillier_accounting_2021} quantify the variability contributed by data sampling, parameter initialization, and hyperparameter selection, and argue that a robust comparison therefore requires multiple trials that account for these sources. Beyond individual experiments, Zhao \etal \cite{10855627} identify recurring weaknesses in foundation-model leaderboard records and documentation that can undermine transparent evaluation and trustworthy model comparison. The common finding is that reproducibility is a property of the pipeline as a whole, not of the model code alone.

A complementary line of work defines reporting and release practices for computational reproducibility. Heil \etal \cite{Heil2021} introduce tiered levels of computational reproducibility (bronze/silver/gold), with requirements covering public code, data, and models, one-command dependency installation, documented execution conditions, deterministic settings, and workflow automation. At the community level, Pineau \etal \cite{Pineau2021} describe the NeurIPS reproducibility program, including a checklist, code policy, and community challenge intended to improve transparency and reproducibility practices across submissions.

Despite these advances, practical support in general ML platforms remains uneven. Gundersen \etal \cite{gundersen2021do} found that none of the evaluated ML platforms supported the full feature set of their proposed reproducibility framework, and observed statistically significant differences when the same experiment was run across four of them. Isdahl and Gundersen \cite{isdahl_out---box_2019} report a comparable gap in a broader survey of out-of-the-box reproducibility support. Both point to the same shortfall: the mechanisms for repeatable execution exist, but they are optional additions rather than defaults of the workflow. Both also grade platforms as general-purpose tools, whereas the comparison in Section~\ref{subsec:methodology-qualitative} grades them against requirements derived from a single application domain and adds direct codebase-change and runtime measurements against two of them.

\subsection{General-Purpose Experimentation Platforms}
General-purpose platforms differ in how users express experiments and how much of the workflow they provide. Graphical tools such as WEKA~\cite{frank_weka_workbench_2016} and visual programming environments such as Orange3~\cite{orange_pdf} allow users to assemble data-processing and model-evaluation workflows with little coding. Jupyter notebooks~\cite{granger_jupyter_2021} instead provide a flexible, code-centric environment in which users define the workflow and its execution structure themselves.

Ludwig~\cite{ludwig_pdf} and Kedro~\cite{kedro} provide more explicit structure through different abstractions. Ludwig supplies a declarative workflow for data preprocessing, model construction, training, hyperparameter optimization, and evaluation, primarily for deep-learning applications. Kedro represents computations as nodes connected through pipelines and manages their inputs and outputs through a Data Catalog, while users implement the nodes and pipeline connections in Python. Ludwig therefore provides more of the model-development procedure, whereas Kedro provides a more general structure for assembling project-specific pipelines.

Between them these platforms cover the graphical, notebook-based, declarative, and structured-pipeline styles of expressing an experiment, which is why all five appear in the qualitative comparison of Section~\ref{subsec:methodology-qualitative}. Jupyter and Kedro serve additionally as the code-centric and structured general-purpose baselines in the quantitative evaluation.

\subsection{Experiment Management and Supporting Infrastructure}
Experiment-management tools primarily record and organize runs performed by user-provided experiment code. MLflow provides experiment tracking together with project packaging, model management, and deployment functionality~\cite{zaharia2018mlflow}. It can be used locally for individual experimentation or deployed as shared infrastructure for teams. MLXP~\cite{arbel_mlxp_2024} is a lightweight, research-oriented system that combines hierarchical configuration, experiment launching, logging, code and job versioning, and result analysis. MLXP uses configuration to launch and manage Python experiment jobs, whereas the proposed design uses configuration to declare a complete application-specific pipeline with explicit stages, dependencies, inputs, and outputs.

MLOps has a broader scope that includes deploying, monitoring, and maintaining ML systems~\cite{kreuzberger2023mlops}. Metaflow supports versioned artifacts and execution from local development to scalable infrastructure~\cite{tagliabue2023metaflow}, Kubeflow Pipelines executes containerized workflows on Kubernetes~\cite{kubeflow_pipelines}, and ZenML connects pipelines to interchangeable infrastructure and MLOps services~\cite{zenml_stacks}. These systems can host research experiments, but their operational capabilities carry infrastructure and maintenance obligations that the lightweight, local workflows targeted here do not warrant.

DVC~\cite{dvc} occupies a narrower role: it provides Git-integrated data and artifact versioning together with declarative, dependency-aware pipelines. It is not a complete application-specific experimentation environment, but supplies execution and versioning mechanisms used by LOCALIZE, in line with FAIR-oriented dataset versioning practice \cite{gonzalezcebrian_standardised_2024}.

\subsection{Application-Specific Research Software}
Application-specific research software packages recurring data-processing procedures, model components, and evaluation methods for a defined class of studies. MONAI~\cite{cardoso_monai_2022} extends PyTorch with transformations, model architectures, and utilities designed for healthcare and medical imaging. MvTS~\cite{ye2024mvts} standardizes data processing, model implementation, training, and evaluation for multivariate time-series forecasting. Both show that software tailored to one application area can absorb the repeated implementation work without cutting its users off from the general ML libraries underneath.

The same principle appears in localization-specific research software. Tong \etal \cite{Tong2021} automate the generation and updating of a theoretical fingerprint database to reduce site-survey effort in CSI localization. This addresses a recurring localization task, but not the complete experimental workflow.

The mechanisms for reproducible experimentation and for reusable domain software therefore exist, but in separate systems, and assembling them is left to each research group. The configuration-first design proposed here draws configuration, stage execution, artifact versioning, and a shared evaluation layer into a single workflow for complete, application-specific experiments. LOCALIZE makes that workflow concrete for radio localization by supplying the corresponding datasets, processing stages, model-development procedures, and evaluation procedures.

\section{Problem statement}
\label{sec:problem}
Existing systems supply most of the mechanisms an ML experiment needs, but they supply them separately: each research group still has to choose them, connect them, and implement the procedures its own studies require. We therefore propose a configuration-first design for application-specific ML experimentation frameworks, instantiate it as LOCALIZE, and evaluate it on radio-localization workflows.

The design pursues three goals. \textit{Reproducibility by default} demands more than fixed seeds: the framework must control the execution context around each stage, so that the only nondeterminism a user has to reason about is the nondeterminism their own code introduces. \textit{Ease of use} means that a complete experiment, from data processing through training to evaluation, is driven by configuration rather than by ad-hoc scripts, and that routine changes require no new code. This matters most where, as in much localization research, new models are evaluated on public datasets and setup work is pure overhead. \textit{Stage modularity} means that a change to one part of an experiment cannot alter an unrelated part through shared state or a hidden dependency.

From these goals we derive five design requirements, stated in Table~\ref{tab:feature-defs}. The same statements serve in Section~\ref{subsec:methodology-qualitative} as the criteria for comparing experimentation platforms, so each is worded to apply to any such platform rather than to LOCALIZE alone, and is paired with the anchors used for grading.

\begin{table*}[t]
    \centering
    \small
    \caption{Design requirements Req.1--Req.5 derived from the three goals of Section~\ref{sec:problem}, with the anchors used when grading platforms against them in Section~\ref{subsec:methodology-qualitative}}
    \label{tab:feature-defs}
    \begin{tabularx}{\textwidth}{@{}p{0.135\textwidth}X p{0.195\textwidth} p{0.195\textwidth}@{}}
        \toprule
        \textbf{Requirement} & \textbf{Definition} & \textbf{Graded \emph{"Very high"}} & \textbf{Graded \emph{"Very low"}} \\
        \midrule

        \textbf{Req.1}\newline Low-code operation &
        A complete experiment, covering data processing, model selection, training, and evaluation, is defined and modified through configuration or a GUI rather than code, which removes boilerplate and lowers the entry barrier. &
        little or no coding &
        substantial coding \\
        \addlinespace[3pt]

        \textbf{Req.2}\newline Reproducibility by default &
        Mechanisms for repeatable execution belong to the default workflow: version control over code, configurations, and data; environment capture; randomness control; isolated execution; and automatic recording of metrics and artifacts. Within one controlled environment, deterministic stages then yield identical recorded results from identical inputs and settings, while custom stages remain responsible for nondeterminism in their own code. &
        integrated mechanisms that need little additional user setup &
        repeatability rests on manual practice and external tooling \\
        \addlinespace[3pt]

        \textbf{Req.3}\newline Common evaluation layer &
        The same metrics, reports, and artifact schema apply across methods and datasets, without bespoke glue code, so that results are directly comparable and interpreted consistently. &
        standardized evaluation that works across models and datasets out of the box &
        no shared evaluation \\
        \addlinespace[3pt]

        \textbf{Req.4}\newline Stage modularity &
        The workflow is decomposed into self-contained stages that communicate through explicitly declared inputs and outputs rather than shared in-memory state, so a stage can be changed or replaced without touching unrelated stages, provided its interface holds. &
        stages have explicit interfaces and exchange state only through them &
        implicit dependencies or shared state, with no stage abstraction \\
        \addlinespace[3pt]

        \textbf{Req.5}\newline Fast setup and iteration &
        Time-to-first-run and time-to-modify are both short: setup is not repeated for every experiment, and a small change does not force a full re-execution, without making the choices involved less explicit or less reproducible. &
        minimal initial setup and efficient iteration, without repetitive reconfiguration or unnecessary reruns &
        heavy installation burden, long configuration times, and slow iteration \\
        \bottomrule
    \end{tabularx}
\end{table*}

\section{Proposed framework}
\label{sec:framework}

The configuration-first design (RQ1) rests on three elements: explicit experiment definitions, isolated stages, and versioned artifacts. It fixes how experiments are configured, executed, and recorded, and leaves each implementation to supply the datasets, transformations, models, and metrics its research domain requires.

\begin{figure}[t]
    \centering
    \includegraphics[width=\columnwidth]{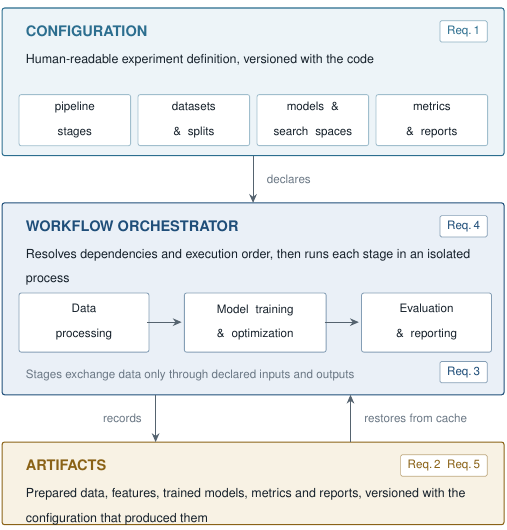}
    \caption{Outline of the proposed architecture consisting of three components: configuration, artifacts, and a workflow orchestrator. Tags mark the design requirement (Req.1--Req.5) that each part addresses}
    \label{fig:architecture}
\end{figure}

\subsection{Design overview}
Fig.~\ref{fig:architecture} shows the three cooperating parts. \textit{Configuration} files describe an experiment in human-readable form, listing its datasets, stages, models, training settings, and evaluation criteria, so that running or altering one is a matter of editing a file rather than writing code.
A \textit{workflow orchestrator} reads that configuration, resolves the dependencies between stages, and executes the pipeline from data processing through reporting, applying the same procedures to every method and keeping runs isolated from one another. An \textit{artifacts} subsystem records the declared inputs, parameters, and outputs at each stage boundary and makes them available for reuse.

The requirements of Section~\ref{sec:problem} map onto these parts as follows. Low-code operation (Req.1) comes from the configuration. Reproducibility by default (Req.2) comes from versioned configurations, isolated execution, and recorded artifacts acting together. A common evaluation layer (Req.3) comes from applying one evaluation procedure to every method. Stage modularity (Req.4) comes from isolating stages behind declared inputs and outputs. Fast iteration (Req.5) comes from cache-aware execution. An application-specific implementation can reduce initial setup further by shipping reusable datasets and templates.

\paragraph{Configuration}
Configuration in Fig.~\ref{fig:architecture} is the primary interface to the framework. Users specify stages, datasets, models, search spaces, evaluation metrics, and other settings within the experiment configuration. The configuration also lists how stages are connected and which implementations they use, so users can follow what will run without inspecting the orchestrator. Because the configuration is itself a file, a change to an experiment takes the form of a diff: it can be reviewed, shared, and version-controlled like source code, and the exact settings behind a result stay attached to that result.

\paragraph{Workflow orchestrator}
The workflow orchestrator in Fig.~\ref{fig:architecture} executes stages according to the configuration. It first resolves the declared dependencies to fix an execution order. It then runs each stage, for every required combination of inputs and parameters, in an isolated process, so that internal state cannot carry from one stage into the next, and routes each stage's outputs to those downstream. Together with the \textit{Artifacts} component of Fig.~\ref{fig:architecture}, it tracks the inputs, parameters, and outputs of every execution, which lets it restore the outputs of an unchanged stage from cache instead of recomputing them and so shortens the iteration cycle. The orchestrator remains stage-agnostic: it implements no stage logic and simply executes the configured plan. Since stages exchange data solely through declared inputs and outputs, a stage can be modified or replaced without touching the orchestrator or any unrelated stage, provided its interface is unchanged.

\paragraph{Reproducibility}
The framework narrows the places where accidental nondeterminism can arise, by fixing the execution order, dependencies, isolation, and versioned state that surround each stage. The supplied components are designed for repeatable execution within a controlled environment. Custom components may introduce new sources of nondeterminism, but explicit inputs, outputs, execution order, and versioned state make their effects easier to isolate and trace.

\paragraph{Experiment pipeline design}
Splitting the workflow into \textit{data processing}, \textit{model training and optimization}, and \textit{evaluation and reporting} draws the boundary where the logic changes character: dataset-specific work falls on one side, method-agnostic work on the other. Preprocessing, models, search spaces, and reporting can then each be changed on their own, and every method is scored by the same procedure.

\textit{Data processing} in Fig.~\ref{fig:architecture} turns raw inputs into an analysis-ready dataset: it parses and cleans the records, optionally constructs features, and materializes the splits that later stages reuse. Confining these operations to a single phase concentrates the dataset-specific logic in one place, so that replacing a dataset means rewriting that phase and nothing else.

\textit{Model training and optimization} selects model hyperparameters from the search spaces declared in the configuration. It consumes the prepared data and the materialized splits, so every method is trained and validated on identical partitions, and adding a model or a hyperparameter combination is a configuration change rather than a pipeline change.

\textit{Evaluation and reporting} aggregates predictions across folds and models, computes a fixed set of metrics, and generates the figures and reports in which the results are presented. Because every method passes through this one stage, differences between reported results reflect the methods rather than the scoring code.

\paragraph{Artifacts}
Artifacts in Fig.~\ref{fig:architecture} capture the intermediate and final results of each experiment: prepared datasets and features, trained model versions, metrics, and reports. Artifacts are versioned together with the configuration that produced them. An earlier result can therefore be revisited, and the state that produced it recovered, without re-implementing the steps that led to it.

\paragraph{Preconfigured datasets}
An application-specific implementation can reduce initial setup by packaging versioned datasets and experiment templates that standardize data access and baseline pipelines.

\begin{figure}[t]
    \centering
    \includegraphics[width=\columnwidth]{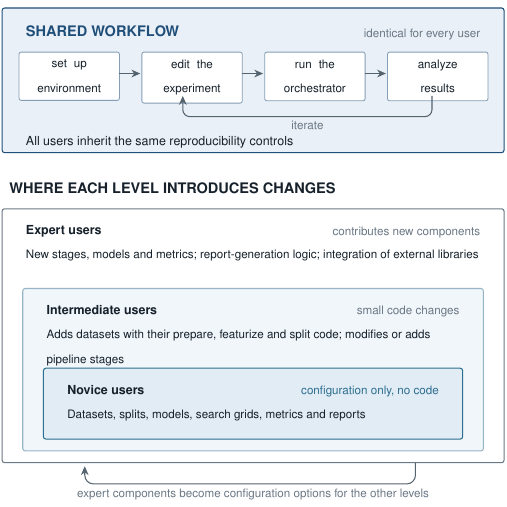}
    \caption{User workflow supported by the framework, separated by experience level. Novice users modify configurations, while intermediate users adapt experiments or add datasets, and expert users implement new stages or models. Nesting indicates that each level keeps the capabilities of the levels it encloses, and components contributed by expert users become configuration options for the others}
    \label{fig:workflow}
\end{figure}

\subsection{User types}
The same workflow accommodates different interaction styles, summarized in Fig.~\ref{fig:workflow}. A routine experiment is configured without writing code; a more demanding one may require changing the pipeline structure or implementing new stage logic. What separates the three levels below is only where a change is introduced, since all of them execute under the same reproducibility controls.

\textit{Novice users} in Fig.~\ref{fig:workflow} run end-to-end experiments by editing configuration files and invoking the orchestrator, without writing code: switching datasets or splits, choosing models and hyperparameter grids, selecting from the available metrics, and configuring reports are all configuration edits.

\textit{Intermediate users} remain configuration-centric but write code where configuration cannot reach, either adding a dataset together with its preparation, featurization, and splitting code, or modifying and adding pipeline stages.

\textit{Expert users} add capabilities: new stages, models, or metrics, integrations with external libraries, or report-generation logic. Once such a component is exposed through the configuration, the other two levels can use it with no further implementation work.

The three modes are complementary rather than alternative: experts contribute reusable components, intermediate users compose them into experiments, and novice users vary those experiments through configuration.

\section{Reference implementation}
\label{sec:implementation}
This section describes how LOCALIZE instantiates the design for a radio localization use case (RQ1) and the alternatives we considered for each component.

\subsection{Backbone: orchestration, versioning, and configuration}
Workflow schedulers such as Airflow~\cite{airflow}, Prefect~\cite{prefect}, and Luigi~\cite{luigi} are well suited to production jobs, but would introduce operational overhead beyond the needs of the local research runs targeted here. Pipeline frameworks such as Snakemake~\cite{snakemake} and Kedro~\cite{kedro} also provide suitable abstractions for research pipelines. We selected DVC~\cite{dvc} paired with Git\footnote{\url{https://git-scm.com}} because DVC records stage commands, dependencies, parameters, and outputs, provides cache-aware execution, and integrates data and artifact versions with source and configuration history. Running \texttt{dvc repro} executes the experiment as declared, while unchanged stages reuse their cached outputs. Each stage command runs as a separate process, providing isolation between stages.

Each experiment directory contains a pipeline stage configuration \texttt{dvc.yaml} that declares the command, dependencies, parameters, inputs, and outputs of every stage. These declarations form the dependency graph that determines what runs and in which order. A separate training configuration \texttt{params.yaml} records dataset splits, models or model architectures, hyperparameter search spaces, evaluation metrics, and other settings. Both \texttt{yaml} files correspond to the Configuration block in Fig.~\ref{fig:architecture}. Because the stage commands reference their source files directly, users can follow the configuration to the code that will execute. Git records the corresponding source and configuration history.

\subsection{Language and environment}
We considered Python, R, Java, and MATLAB for the core runtime. Although MATLAB still appears in localization work \cite{matlab_survey}, recent studies and toolchains lean heavily on Python \cite{Roy2021MLSurvey, python_survey}. We chose Python for its mature ML ecosystem, including NumPy, Pandas, scikit-learn, TensorFlow, and Keras. The framework is distributed as a standard Python repository. Users clone it and create the environment from an \texttt{environment.yaml} file using Conda or Mamba.

\subsection{Implementing the three-stage design}
\begin{figure}[t]
    \centering
    \includegraphics[width=\columnwidth]{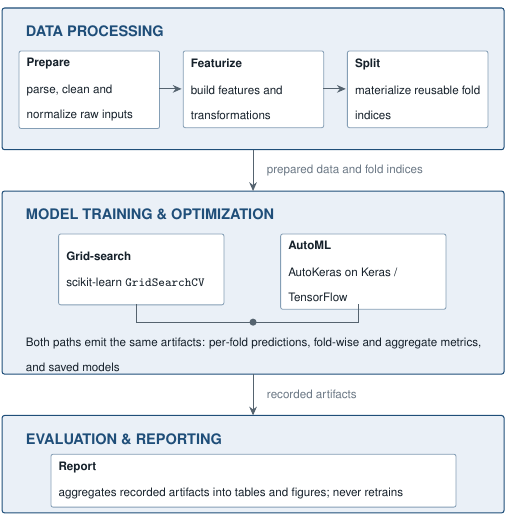}
    \caption{Decomposition of the three conceptual phases of Fig.~\ref{fig:architecture} into the executable stages supplied by LOCALIZE. Each stage runs as a separate process, with its dependencies, parameters, inputs and outputs declared in \texttt{dvc.yaml} and \texttt{params.yaml}}
    \label{fig:stages}
\end{figure}

Conceptually, the framework has three stages as depicted in Fig.~\ref{fig:architecture}: \textit{data processing}, \textit{model training and optimization}, and \textit{evaluation and reporting}. LOCALIZE divides each of them into narrower executable units, aligned with the methodology behind the knowledge discovery process \cite{10.1145/240455.240464}, so that a change touches as little of the pipeline as possible and each unit corresponds to the library that performs the work. The units exchange data through declared file dependencies and outputs. Fig.~\ref{fig:stages} shows the resulting decomposition.

\paragraph{Data processing}
We implement data processing from Fig. \ref{fig:architecture} as three stages to keep concerns separate and caching effective. First, the \emph{Prepare} stage parses raw inputs, cleans malformed records, and normalizes the schema into a consistent form. Keeping parsing independent prevents feature edits from re-triggering heavy I/O work. Second, the
\emph{Featurize} stage constructs features and transformations. Feature sets can then be iterated on without touching raw-data handling, and without invalidating the cached output of Prepare. Third, the \emph{Split} stage materializes the cross-validation strategy as explicit fold indices that are reused across models. This keeps the data partitions identical across models and reruns.

\paragraph{Model training and optimization}
The design has one training stage in Fig. \ref{fig:architecture}, but classical and neural pipelines use different runtimes and produce different artifacts. We therefore provide two executable paths and standardize their outputs. \emph{Grid-search} is implemented with scikit-learn\footnote{\url{https://scikit-learn.org}} using \texttt{GridSearchCV}. It offers a stable API across many algorithms, integrates naturally with NumPy/Pandas, accepts user-supplied fold indices, and can produce repeatable results when the estimator, data splits, random seeds, and execution environment are controlled. Explicit grids map cleanly to YAML and are practical for the small classical-ML search spaces used here. For larger or more complex search spaces, LOCALIZE provides a separate AutoML path.

The \emph{AutoML} path is implemented with AutoKeras\footnote{\url{https://autokeras.com}} on top of Keras/TensorFlow. It maps cleanly onto a configuration-first workflow and allows experts to define new blocks or expand search spaces using Keras and KerasTuner without changing the surrounding pipeline.

Both paths produce a common set of downstream artifacts: per-fold predictions, fold-wise and aggregate metrics, saved model artifacts in pickle or Keras format, and a compact run summary. Scoring and reporting are therefore identical across methods, and switching between classical and neural models leaves the reporting stage untouched.
\paragraph{Evaluation and reporting}
Evaluation runs inside each training backend because data structures and APIs differ. Outputs are then normalized into this common artifact structure. A separate \textit{Report} stage aggregates results across runs and methods into tables and figures, which are recorded as artifacts in Fig. \ref{fig:architecture}. Reporting uses only recorded artifacts and does not retrain models, allowing reports to be modified without rerunning the training stages.
\subsection{Reproducibility}
In LOCALIZE, Git versions the source code, configurations, and environment specification, while DVC records the dependencies, parameters, and outputs needed to reproduce each pipeline stage. Each DVC stage is executed in a separate process, preventing in-memory state from carrying between stages. Materialized splits and controlled random seeds further address model-level variation. Together, these mechanisms make the supplied pipelines repeatable from identical inputs and settings within the same controlled environment; custom stages remain responsible for controlling nondeterminism introduced by their own code.

\subsection{Preconfigured datasets}
As part of artifacts in Fig. \ref{fig:architecture}, the repository includes five preconfigured datasets, dataset-specific Prepare and Featurize stages, reusable Split stages for several cross-validation strategies, and experiment templates. These resources cover the Lumos5G dataset \cite{narayanan_lumos5g_2020} and the UMU dataset \cite{asensio-garriga_university_2025} for cellular localization; the LOG-a-TEC dataset \cite{morano_outdoor_2022} for Bluetooth Low Energy localization; and the CTW 2019 \cite{arnold_novel_2019} and CTW 2020 \cite{gauger_massive_2020} datasets, used here for CSI-based localization from massive-MIMO channel measurements. To experiment with one of these datasets, users copy a template, edit \texttt{params.yaml}, and run the study without writing dataset-integration code or defining the pipeline configuration from scratch.
\section{Evaluation methodology}
\label{sec:methodology}

We evaluate the design of Section~\ref{sec:framework}, as instantiated by LOCALIZE in Section~\ref{sec:implementation}, through four studies: a cross-platform qualitative comparison (RQ1), a controlled lines-of-code study (RQ2), stage-level runtime and resource profiling (RQ3), and a dataset-volume scaling study (RQ4).

\subsection{Cross-platform qualitative comparison (RQ1)}
\label{subsec:methodology-qualitative}

As the first step in the evaluation of LOCALIZE, we selected a representative rather than exhaustive set of comparison platforms: WEKA~\cite{frank_weka_workbench_2016}, Orange3~\cite{orange_pdf}, Ludwig~\cite{ludwig_pdf}, Jupyter notebooks in JupyterLab~\cite{granger_jupyter_2021}, and Kedro~\cite{kedro}. Together, they cover graphical, notebook-based, configuration-driven, and structured-pipeline approaches to local, end-to-end ML experimentation. Other systems discussed in Section~\ref{sec:related} were not graded individually because they primarily provide supporting experiment-tracking or infrastructure capabilities, target scalable or production operation, or address application-specific workflows outside the scope of this evaluation. DVC was not assessed separately because it forms LOCALIZE's execution and versioning layer. Among general-purpose pipeline frameworks, Kedro was selected because its local nodes, data dependencies, parameters, and Data Catalog map closely to the controlled workflow.

To answer RQ1, we grade each platform against Req.1--Req.5 as stated in Table~\ref{tab:feature-defs}, considering the localization workflows targeted by LOCALIZE. Table~\ref{tab:qualitative-comparison} additionally records a descriptive \emph{Primary focus} for each platform, which is not graded. The criteria accommodate GUI-first, configuration-driven, and code-centric platforms through their normal interaction models. We assign ordinal grades based on publicly available official documentation and hands-on use. The grades indicate suitability for the targeted localization workflows rather than the platforms' general quality within their intended domains. The assessment includes capabilities available through each platform's documented installation and normal workflow, including directly integrated dependencies. If a capability requires a separately selected and configured tool, its grade is capped at \emph{Medium}.

\subsection{Materials}

The quantitative experiments rely on the UMU cellular localization dataset~\cite{asensio-garriga_university_2025} and the LOG-a-TEC Bluetooth localization dataset~\cite{morano_outdoor_2022}. The initial pipeline in the LOC study used UMU, while Change 2 and the subsequent changes, as well as the runtime and dataset-volume scaling studies, used LOG-a-TEC.

All pipeline executions for the quantitative evaluation ran on a Kubernetes cluster with two AMD EPYC~75F3 CPUs, providing 128 hardware threads in total, and 1\,TB of RAM. Each execution ran inside an isolated JupyterLab container with access to the full CPU and memory resources; the profiling protocol separately imposed the workload-level CPU limits described in Section~\ref{subsec:methodology-resource}. The software stack comprised Python~3.11.13, scikit-learn~1.4.2, TensorFlow~2.16.2, Keras~3.5.0, AutoKeras~2.0.0, DVC~3.67.1, Kedro~1.5.0, kedro-datasets~9.5.0, Git, Mamba, and psutil~7.2.2. The three implementations used separate environment specifications with identical constraints for their shared dependencies. All computations used the CPU only.

\subsection{Lines of code study}
\label{subsec:methodology-loc}

The qualitative comparison covers six platforms, while the quantitative studies use Jupyter and Kedro as baselines. The controlled LOC study (RQ2) does not translate to GUI-first tools such as WEKA and Orange3, and reproducing the same stage boundaries and instrumentation in those environments would not yield a like-for-like resource comparison; Ludwig was excluded because it primarily targets deep-learning workflows. The two retained baselines bracket the design space: Jupyter is a flexible code-centric workflow with little imposed structure, and Kedro a structured, general-purpose pipeline framework.

To answer RQ2, we implemented corresponding pipelines in LOCALIZE, a Jupyter notebook, and Kedro. All three followed the same sequence of experimental changes and used the same models, datasets, splitting strategies, and evaluation settings at each version, while retaining the normal structure of each implementation. The initial pipelines were composed of the stages \emph{Prepare}, \emph{Featurize}, \emph{Split}, and \emph{Grid-search} for the UMU dataset. These pipelines initially used the \texttt{LinearRegression} model from scikit-learn, a two-value hyperparameter grid for \texttt{fit\_intercept}, 5\nobreakdash-fold \texttt{KFold}, and RMSE as the evaluation metric. We then introduced four incremental changes:
\textit{Change 1} replace the estimator and add a second one,
\textit{Change 2} switch the dataset from UMU to LOG-a-TEC,
\textit{Change 3} add a second data-splitting strategy and an additional evaluation metric,
\textit{Change 4} append an AutoML stage using AutoKeras with 10 trials and a maximum of 1000 epochs.

For each change, we counted \emph{lines of code (LOC) added} and \emph{LOC removed} relative to the preceding version and reported them separately. For LOCALIZE, the count includes \texttt{dvc.yaml} and \texttt{params.yaml}. For Jupyter, the count includes executable notebook cells; for Kedro, it includes \texttt{catalog.yml}, \texttt{parameters.yml}, and the node and pipeline source code. Shared setup, environment, and benchmarking code was excluded because it did not form part of the controlled changes. We also excluded changes that only renamed experiment-specific output paths. Blank lines, comments, and lines containing only structural tokens (\eg, brackets, braces, or commas) were not counted.

\subsection{Runtime and resource profiling}
\label{subsec:methodology-resource}

To answer RQ3, we profiled the final LOCALIZE, Jupyter, and Kedro pipelines from the LOC study to compare the runtime and resource usage of individual pipeline stages. We measured five stages: \emph{Prepare}, \emph{Featurize}, \emph{Split}, \emph{Grid-search}, and \emph{AutoML}. Each workload process and its descendants were restricted to five logical CPUs, while the resource-monitoring process was assigned an additional logical CPU. Measurements covered the loading, execution, and output handling of each measured stage. LOCALIZE starts each stage in a separate process, while the Jupyter kernel and Kedro process remain active across stages. The LOCALIZE boundary includes stage-process startup, while Jupyter kernel startup and Kedro project initialization before the first catalog load are excluded. This difference mainly affects the short stages.

Memory and CPU usage were sampled at 10\,Hz using \texttt{psutil} and aggregated across the workload process and its descendants. CPU time, expressed in core-seconds, was obtained by integrating the time-series CPU usage data using the trapezoidal rule. Wall time was measured using a high-resolution performance counter. Memory usage was measured as proportional set size (PSS).

Each pipeline was executed five times under the same configuration. For stages consisting of multiple jobs, CPU and wall times were summed within each run. For CPU and wall time, the figures report the mean, first and third quartiles, and observed maximum across the five runs. Memory samples from all five runs were combined to calculate the same statistics. Complete-pipeline wall time was calculated by summing the stage wall times within each run and then averaging the resulting totals across the five runs. Complete-pipeline peak memory was the largest PSS value observed across all stages and runs.

\subsection{Scaling with increased data volume}
\label{subsec:methodology-logatec}
To examine how resource usage changes with dataset volume and answer RQ4, we reused the final LOCALIZE pipeline from the LOC study and increased the raw dataset volume to 1$\times$, 5$\times$, and 10$\times$ the base LOG-a-TEC size by repeating the raw measurement records with shifted timestamps. The scaled raw data were then processed by the complete pipeline, including the Prepare stage. Each size was run once. Memory usage, CPU time, and wall time were computed for all five stages, including AutoML, with the same measurement procedure as above. For each dataset size, aggregate CPU and wall time were calculated by summing the corresponding values across the five stages. The overall maximum memory was the largest PSS value observed across the stages.

\subsection{Repeatability and reproducibility measures}
\label{subsec:methodology-repro}
Configurations, code, data versions, environments, and random seeds were controlled throughout the evaluation. Within each implementation, the generated Grid-search and AutoML metric summaries were checked for exact equality across the five repeated runs. Following ACM/NASEM terminology \cite{acm_badging_2020}, this establishes \emph{repeatability} (same team, same setup). \emph{Reproducibility}, meaning a different team obtaining the same result from the released artifacts, is supported by the public code, configurations, and environment specifications.

\begin{table*}[t]
    \centering
    \footnotesize
    \setlength{\tabcolsep}{3pt}
    \renewcommand{\arraystretch}{1.08}
    \caption{Qualitative comparison of LOCALIZE and representative platforms for the targeted localization workflows, using the requirements defined in Table~\ref{tab:feature-defs}. Platform selection is described in Section~\ref{subsec:methodology-qualitative}}
    \label{tab:qualitative-comparison}
    \begin{tabular}{@{}>{\raggedright\arraybackslash}p{0.117\textwidth}
            >{\centering\arraybackslash}p{0.146\textwidth}
            >{\centering\arraybackslash}p{0.11\textwidth}
            >{\centering\arraybackslash}p{0.12\textwidth}
            >{\centering\arraybackslash}p{0.183\textwidth}
            >{\centering\arraybackslash}p{0.114\textwidth}
            >{\centering\arraybackslash}p{0.095\textwidth}@{}}
        \toprule
        \textbf{Item}$^{\mathrm{a}}$ & \textbf{LOCALIZE} & \textbf{WEKA} & \textbf{Orange3}             & \textbf{Jupyter notebook}    & \textbf{Ludwig} & \textbf{Kedro}         \\
        \midrule
        \textbf{Primary focus}       & Localization      & General ML    & Data mining \& visualization & Experimentation \& scripting & Deep learning   & Data science pipelines \\
        \textbf{Req.1}               & Very high         & Very high     & Very high                    & Very low                     & High            & Medium                 \\
        \textbf{Req.2}               & High              & Medium        & Medium                       & Low                          & Medium          & Medium                 \\
        \textbf{Req.3}               & Very high         & Very high     & Very high                    & Very low                     & High            & Very low               \\
        \textbf{Req.4}               & Very high         & High          & High                         & Low                          & Medium          & High                   \\
        \textbf{Req.5}               & Very high         & High          & High                         & High                         & High            & High                   \\
        \bottomrule
    \end{tabular}
    \vspace{2pt}
    \begin{minipage}{\textwidth}
        \scriptsize $^{\mathrm{a}}$\,Req.1 = Low-code operation; Req.2 = Reproducibility by default; Req.3 = Common evaluation layer; Req.4 = Stage modularity; Req.5 = Fast setup and iteration. Full definitions are given in Table~\ref{tab:feature-defs}.
    \end{minipage}
\end{table*}

\section{Evaluation}
\label{sec:evaluation}

We report the three studies in turn: the qualitative comparison (RQ1), the quantitative comparison against Jupyter and Kedro (RQ2 and RQ3), and the LOG-a-TEC scaling experiment (RQ4).
\subsection{Cross-platform qualitative comparison (RQ1)}
Table~\ref{tab:qualitative-comparison} reports the \emph{Primary focus} and the Req.1--Req.5 grades for the six platforms; the paragraphs below give the basis for each grade.
\paragraph{Primary focus (descriptive)}
The first row of Table~\ref{tab:qualitative-comparison} describes the scope of the platforms and is not scored. \textsc{LOCALIZE} targets end-to-end localization workflows, WEKA general classical ML, Orange3 visual data mining, Jupyter code-centric experimentation, Ludwig configuration-driven deep learning, and Kedro modular data science pipelines.

\paragraph{Req.1: Low-code operation}
Three platforms are rated \emph{"Very high"} on Req.1. Within its supplied localization workflows, \textsc{LOCALIZE} allows users to select datasets, splitting strategies, models, hyperparameter grids, metrics, and reports through configuration without writing code. Req.1 concerns this routine use of supported workflows; implementing new components falls under Req.4. WEKA and Orange3 allow users to assemble workflows through graphical interfaces with little or no coding. Jupyter is rated \emph{"Very low"}, since routine changes require writing and maintaining code. Ludwig is rated \emph{"High"}, as its configuration covers preprocessing, model construction, training, hyperparameter optimization, and evaluation, while operations outside its abstractions require custom code. Kedro is rated \emph{"Medium"}: its catalog and parameters are configuration-driven, but pipeline nodes and their connections are defined in Python.

\paragraph{Req.2: Reproducibility by default}
\textsc{LOCALIZE} is rated \emph{"High"} on Req.2, because configuration, environment management, execution isolation, and artifact handling are integrated into the normal workflow. WEKA, Orange3, Ludwig, and Kedro are rated \emph{"Medium"}. WEKA can save experiment definitions and results, Orange3 persists workflows and widget settings, Ludwig records configurations, random seeds, models, training statistics, and evaluation results, and Kedro provides project configuration together with optional dataset and model versioning through its Data Catalog. In each case, however, complete environment capture, deterministic execution controls, and end-to-end versioning remain project-level decisions left to the user. Jupyter is rated \emph{"Low"}, because reproducibility depends on user discipline and external tooling.

\paragraph{Req.3: Common evaluation layer}
\textsc{LOCALIZE}, WEKA, and Orange3 are rated \emph{"Very high"} on Req.3. \textsc{LOCALIZE} applies common metrics, reporting, and artifact schemas across datasets and methods. WEKA Experimenter provides standardized evaluation and result storage across algorithms and datasets, while Orange3 applies common validation procedures and metrics to multiple learners and connects the results to reporting and visualization widgets. Ludwig is rated \emph{"High"} because it automatically produces evaluation metrics, model artifacts, and result files for deep-learning workflows, but does not extend the same evaluation layer to classical ML pipelines. Jupyter and Kedro are rated \emph{"Very low"} because neither provides a built-in shared model-evaluation procedure or result schema; these must be implemented in experiment-specific code.

\paragraph{Req.4: Stage modularity}
On Req.4, \textsc{LOCALIZE} is rated \emph{"Very high"} because its stages exchange data through declared files and run in separate processes. WEKA, Orange3, and Kedro are rated \emph{"High"} because they connect explicit processing steps, widgets, or nodes, but do not isolate component state to the same degree in their default workflows. Ludwig is rated \emph{"Medium"} because its configuration separates parts of the workflow without exposing them as a user-composable stage graph. Jupyter is rated \emph{"Low"} because notebook cells normally share a kernel and depend on implicit execution order.

\paragraph{Req.5: Fast setup and iteration}
\textsc{LOCALIZE} is rated \emph{"Very high"} on Req.5, as it combines ready-to-run datasets and templates with cached, stage-wise execution, so common changes do not require full reruns. WEKA, Orange3, Jupyter, Ludwig, and Kedro are rated \emph{"High"}. They offer relatively direct setup or iteration through graphical workflows, notebooks, configuration, or project templates, but do not combine these features with preconfigured localization datasets and cached stage-wise execution. \textsc{LOCALIZE}'s grade therefore rests on its packaged localization resources as much as on the underlying framework.

\noindent\fbox{%
    \parbox{\dimexpr\linewidth-2\fboxsep-2\fboxrule}{%
        \textit{Answer to RQ1}: Such a platform declares experiments in versioned configuration rather than in scripts (Req.1, Req.2), executes each stage as an isolated process communicating only through declared inputs and outputs (Req.4), records those as versioned artifacts that also drive cache-aware re-execution (Req.5), and scores every method through one evaluation layer built on them (Req.3); none of the five comparison platforms combines all of these properties for the targeted workflows (Table~\ref{tab:qualitative-comparison}).
    }%
}

\begin{figure}[t]
    \centering
    \includegraphics[width=\columnwidth]{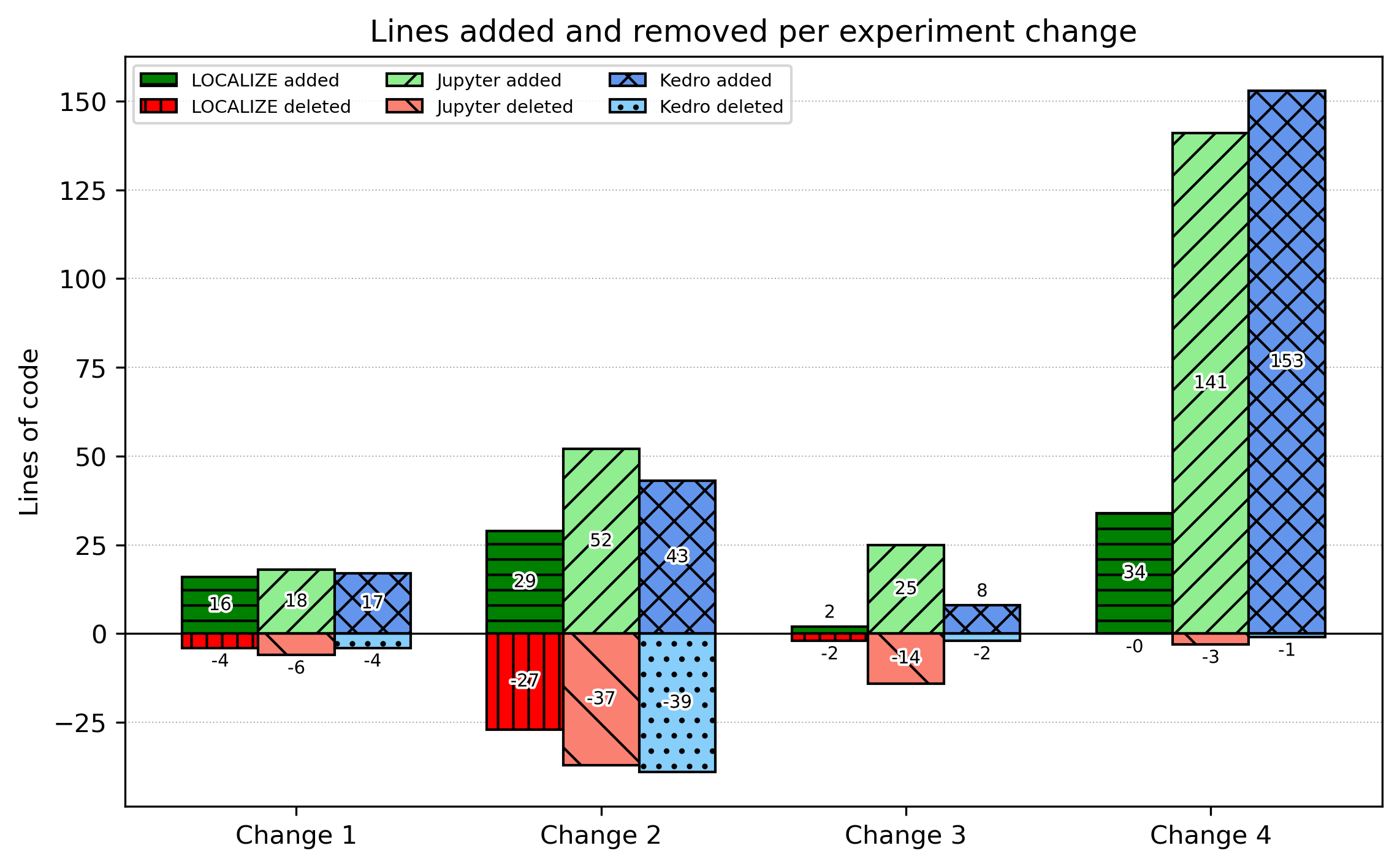}
    \caption{Lines of code added (positive) and removed (negative) when introducing successive changes in the LOCALIZE, Jupyter, and Kedro implementations}
    \label{fig:loc-change-fig}
\end{figure}
\begin{figure*}[t]
    \centering
    \includegraphics[width=0.8\linewidth]{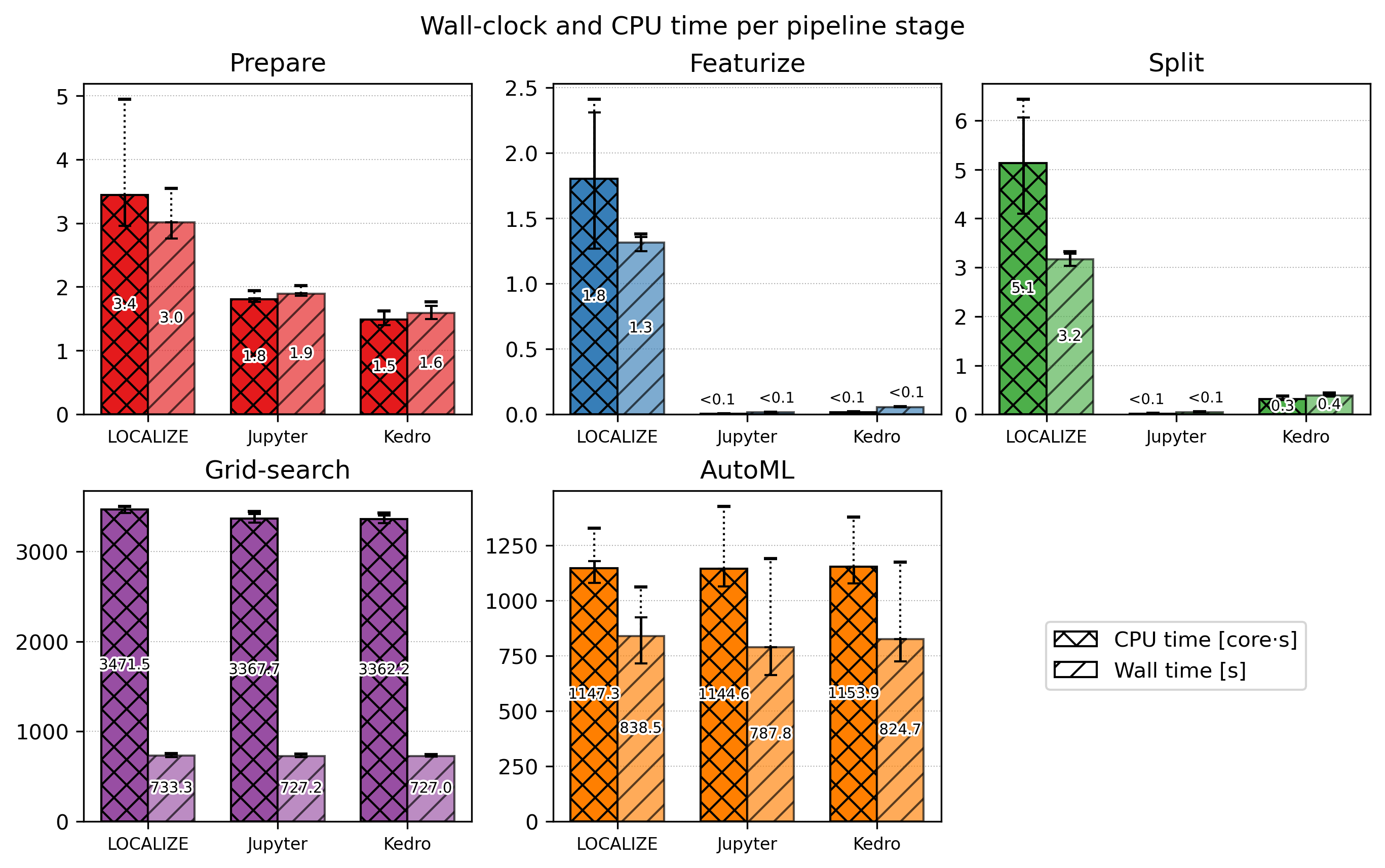}
    \caption{Wall-clock and CPU time per pipeline stage for LOCALIZE, Jupyter, and Kedro. Bars show means, solid error bars show the first and third quartiles, and dotted whiskers show the observed maxima across five runs}
    \label{fig:cpu-wall-time-fig}
\end{figure*}
\begin{figure}[t]
    \centering
    \includegraphics[width=\columnwidth]{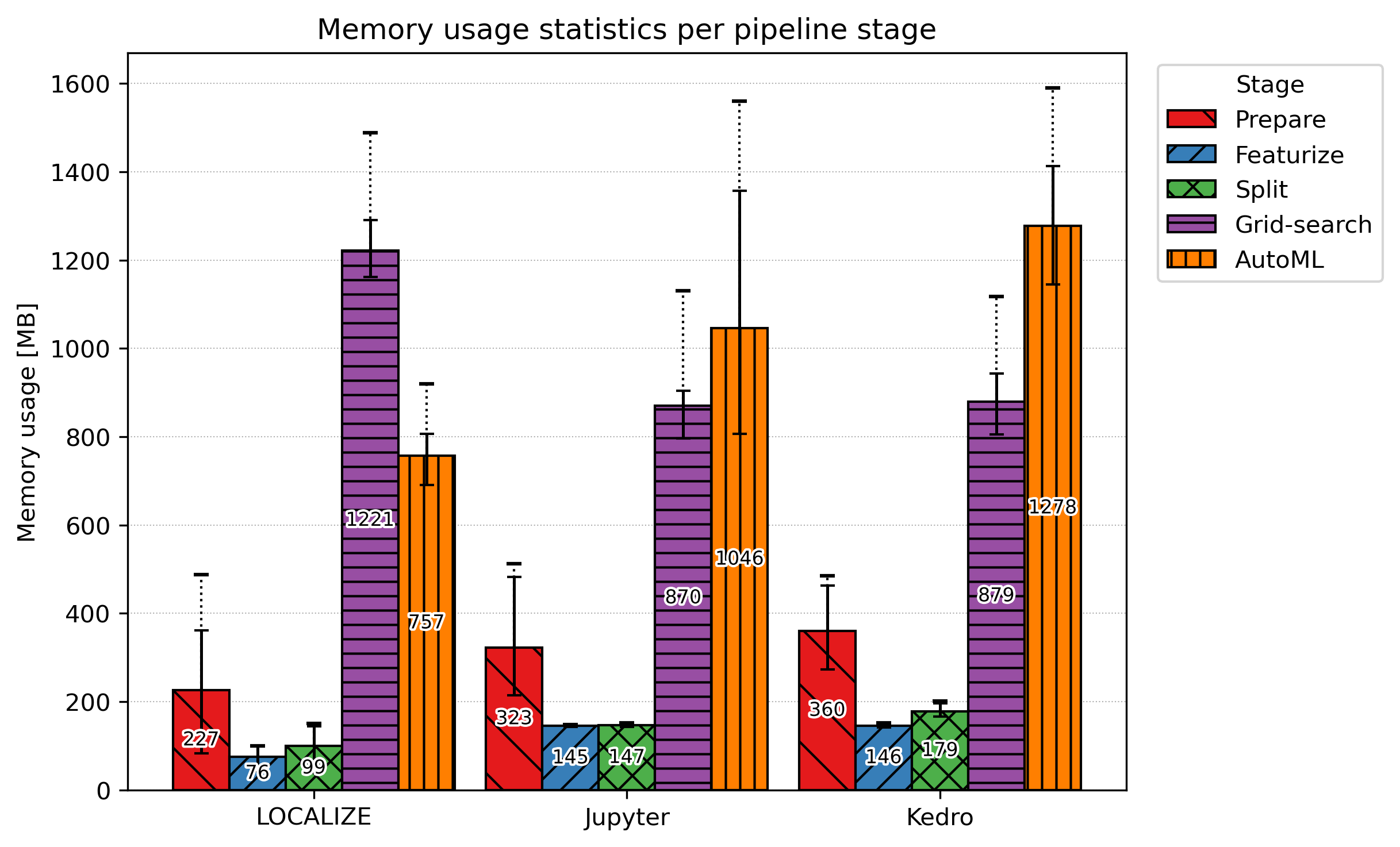}
    \caption{PSS memory usage per pipeline stage for LOCALIZE, Jupyter, and Kedro. Statistics combine samples from all five runs: bars show means, solid error bars show the first and third quartiles, and dotted whiskers show the observed maxima.}
    \label{fig:mem-usage-fig}
\end{figure}

\subsection{Comparison against Jupyter and Kedro baselines}

\paragraph{Lines of code comparison (RQ2)}
Figure~\ref{fig:loc-change-fig} compares the codebase changes required for the four incremental changes in \textsc{LOCALIZE}, Jupyter, and Kedro. Positive bars show lines added and negative bars show lines removed, counted as described in Section~\ref{subsec:methodology-loc}.

Change~1, replacing one estimator and adding another, requires similar changes in all three implementations: +16/-4 LOC in LOCALIZE, +18/-6 in Jupyter, and +17/-4 in Kedro. Change~2, switching from UMU to LOG-a-TEC, requires +29/-27 LOC in LOCALIZE, compared with +52/-37 in Jupyter and +43/-39 in Kedro. For Change~3, LOCALIZE adds a second splitting strategy and metric with +2/-2 LOC, while Jupyter requires +25/-14 and Kedro +8/-2. The largest difference occurs in Change~4. Adding AutoML requires +34/-0 LOC in LOCALIZE, +141/-3 in Jupyter, and +153/-1 in Kedro.

These counts show where each implementation draws the boundary between built-in functionality and experiment-specific code. LOCALIZE expresses the studied changes through configuration wherever an existing component covers them, whereas the Jupyter and Kedro implementations had to carry the AutoML training, validation, scoring, and result-handling logic in the experiment itself. The measure is therefore the size of the codebase change in these implementations, and not development time, cognitive effort, or framework productivity in general.
\noindent\fbox{%
    \parbox{\dimexpr\linewidth-2\fboxsep-2\fboxrule}{%
        \textit{Answer to RQ2}: For changes its supplied components already cover, LOCALIZE required fewer codebase edits than the Jupyter and Kedro implementations, most markedly when adding AutoML (+34/-0 against +141/-3 and +153/-1); replacing an estimator was comparable across all three.
    }%
}

\begin{figure*}[t]
    \centering
    \includegraphics[width=0.8\linewidth]{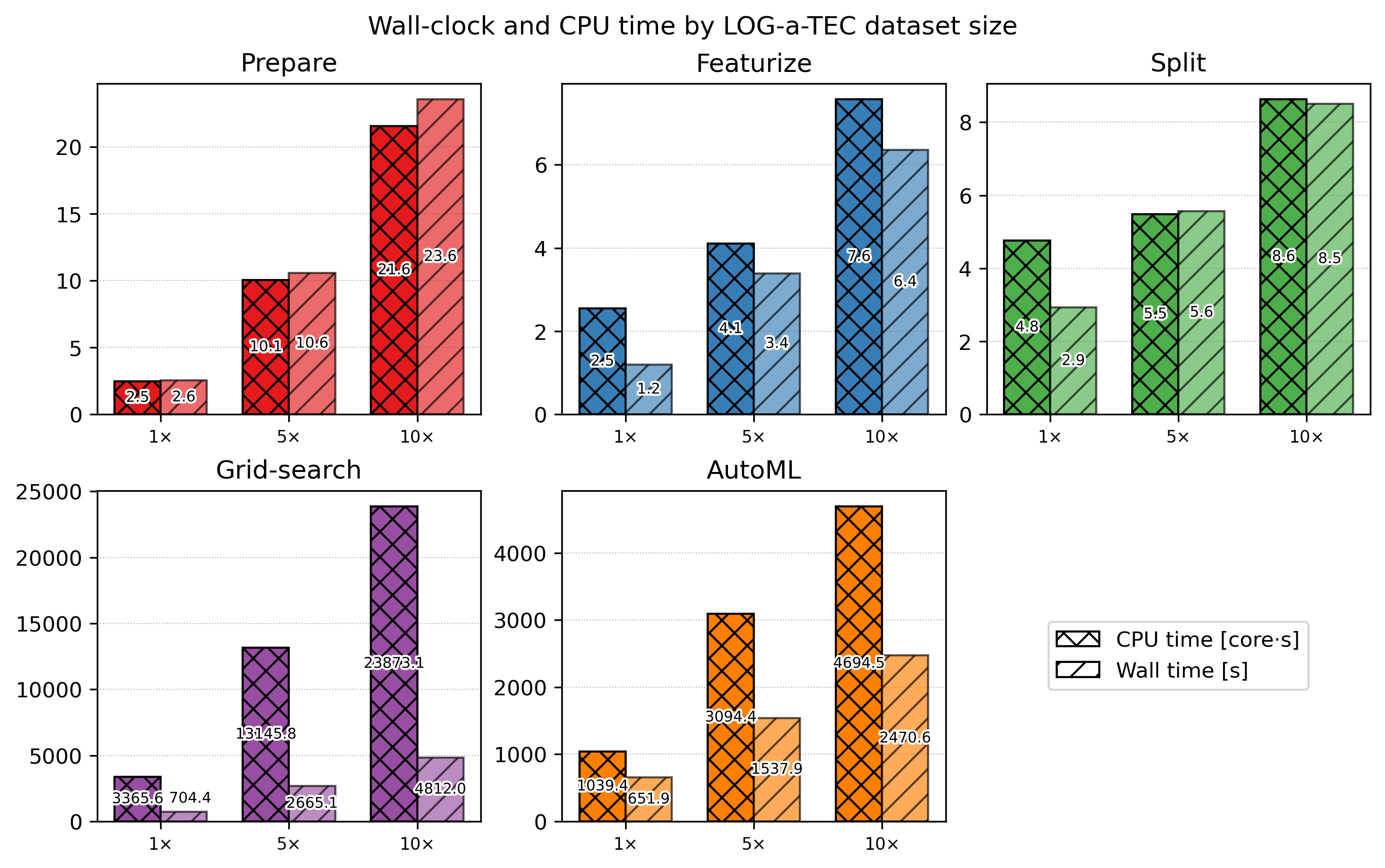}
    \caption{Wall-clock and CPU time per stage for LOCALIZE on 1$\times$, 5$\times$, and 10$\times$ LOG-a-TEC}
    \label{fig:cpu-wall-time-logatec-fig}
\end{figure*}

\paragraph{CPU and wall time by stage (RQ3)}
Figure~\ref{fig:cpu-wall-time-fig} compares CPU and wall time by pipeline stage. LOCALIZE is the more expensive implementation in all three short stages: Prepare costs 3.4 core-seconds, compared with 1.8 for Jupyter and 1.5 for Kedro; Featurize costs 1.8 core-seconds, compared with under 0.1 for both baselines; and Split costs 5.1 core-seconds, compared with under 0.1 for Jupyter and 0.3 for Kedro. Since all three stages are short relative to training, these differences are consistent with the process-startup and measurement-boundary differences described in Section~\ref{subsec:methodology-resource}.

Grid-search is the most CPU-intensive stage. LOCALIZE uses 3472 core-seconds and 733 seconds of wall time, compared with 3368 and 727 seconds for Jupyter and 3362 and 727 seconds for Kedro. LOCALIZE therefore uses approximately 3\% more CPU time and less than 1\% more wall time. LOCALIZE retains models for individual candidates and folds to support its reporting and artifact-tracking functionality, while Jupyter and Kedro use the standard \texttt{GridSearchCV} behavior; despite this additional work, Grid-search wall time remains similar. AutoML is also similar across the implementations: CPU time ranges from 1145 to 1154 core-seconds and wall time from 788 to 839 seconds. Summed across stages, total wall time is 1579 seconds for LOCALIZE, 1517 seconds for Jupyter, and 1554 seconds for Kedro. The observed wall times are therefore comparable across the complete pipelines.

\paragraph{Memory by stage (RQ3)}
Figure~\ref{fig:mem-usage-fig} compares PSS memory usage by pipeline stage. LOCALIZE uses less memory than either baseline in the three short stages and more in Grid-search, where it averages 1221 MB and peaks at 1489 MB against averages of 870 and 879 MB and peaks of 1131 and 1118 MB for Jupyter and Kedro. The excess is consistent with the additional model and artifact retention described above.

The pattern reverses in AutoML. LOCALIZE averages 757 MB, compared with 1046 MB for Jupyter and 1278 MB for Kedro. LOCALIZE starts AutoML in a fresh interpreter, while the other implementations retain process state from earlier stages. These stage-level differences largely cancel over the pipeline: peak memory is 1489 MB for LOCALIZE, 1560 MB for Jupyter, and 1590 MB for Kedro, a spread of roughly 7\%.

\noindent\fbox{%
    \parbox{\dimexpr\linewidth-2\fboxsep-2\fboxrule}{%
        \textit{Answer to RQ3}: Total wall time and pipeline peak memory were comparable across the three implementations; LOCALIZE traded ${\sim}$3\% more Grid-search CPU time and higher Grid-search memory, both from per-fold model retention, against lower AutoML memory and a per-process startup cost in the short stages.
    }%
}

\subsection{Scaling with training data volume}

\paragraph{Per-stage CPU and wall time (RQ4)}
Figure~\ref{fig:cpu-wall-time-logatec-fig} compares per-stage CPU and wall time at the 1$\times$, 5$\times$, and 10$\times$ dataset sizes. Most of the growth occurs in Grid-search and AutoML. Grid-search wall time increases from 704 seconds at 1$\times$ to 2665 seconds at 5$\times$ and 4812 seconds at 10$\times$. AutoML wall time increases from 652 seconds at 1$\times$ to 1538 seconds at 5$\times$ and 2471 seconds at 10$\times$.

\begin{figure}[t]
    \centering
    \includegraphics[width=\columnwidth]{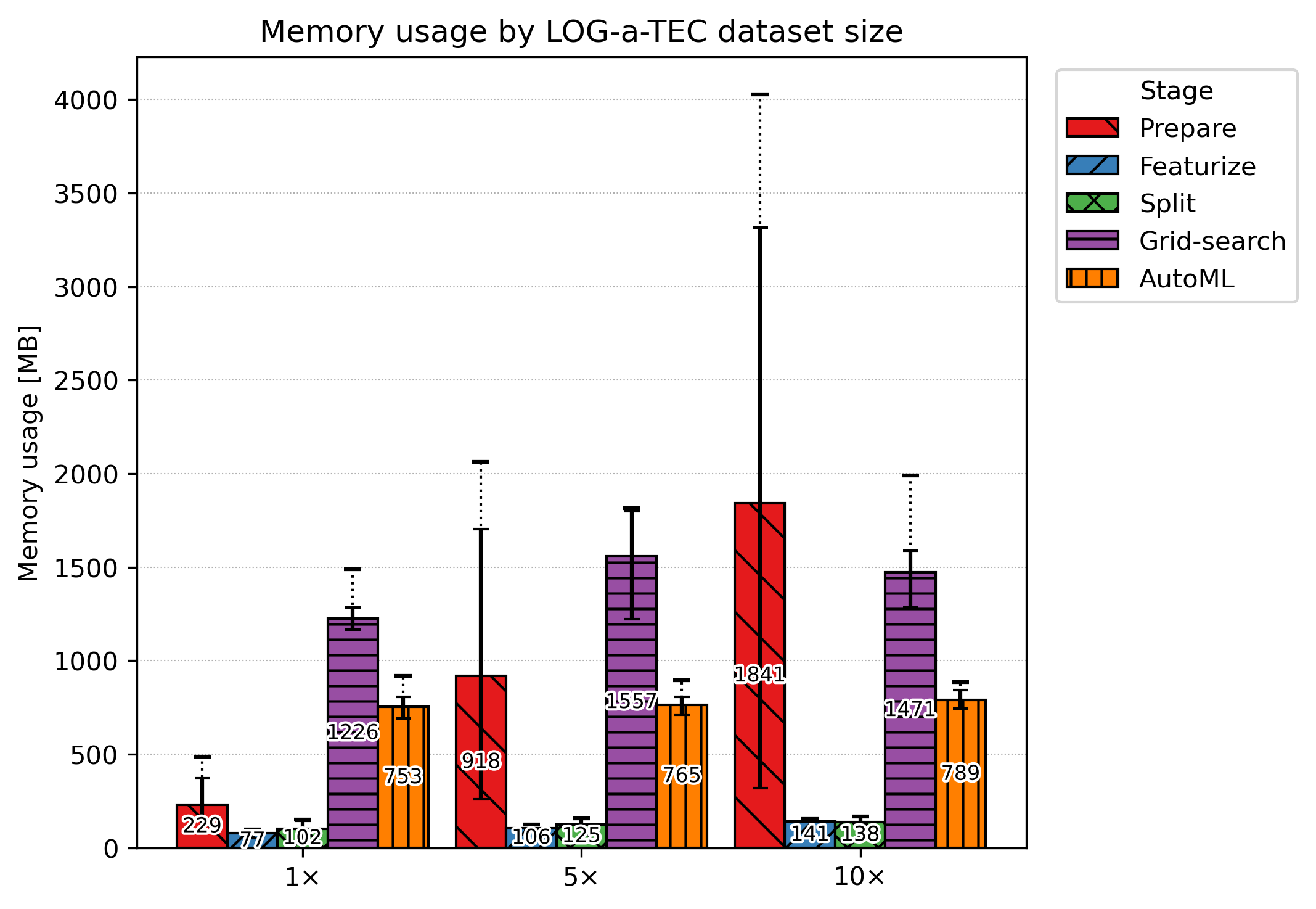}
    \caption{PSS memory usage per stage for LOCALIZE on 1$\times$, 5$\times$, and 10$\times$ LOG-a-TEC. Bars show means, solid error bars show the first and third quartiles, and dotted whiskers show the observed maxima}
    \label{fig:mem-usage-logatec-fig}
\end{figure}

\paragraph{Per-stage memory usage (RQ4)}
Figure~\ref{fig:mem-usage-logatec-fig} shows that memory growth is stage-dependent. Featurize and Split remain below the two training stages at all three dataset sizes. Prepare grows from 229 MB at 1$\times$ to 918 MB at 5$\times$ and 1841 MB at 10$\times$, becoming the largest mean-memory stage at 10$\times$; its observed maximum reaches 4027 MB. AutoML mean memory is essentially flat, rising only from 753 to 789 MB across the three sizes. Grid-search averages 1226, 1557, and 1471 MB, while its observed maximum rises from 1489 to 1814 and 1991 MB. The lower 10$\times$ Grid-search mean relative to 5$\times$ does not indicate a smaller peak memory requirement. The mean is calculated across samples collected during jobs of different durations, and each size was measured once.

\paragraph{Aggregate CPU and wall time (RQ4)}
Table~\ref{tab:cpu-usage-table} aggregates CPU and wall time across all stages.

\begin{table}[htbp]
    \centering
    \caption{CPU and wall time for multiples of the LOG-a-TEC dataset (1$\times$, 5$\times$, and 10$\times$ the base size)}
    \label{tab:cpu-usage-table}
    \begin{tabularx}{0.9\linewidth}{@{}Xccc@{}}
        \toprule
        \textbf{Metric}               & \textbf{1$\times$ base} & \textbf{5$\times$ base} & \textbf{10$\times$ base} \\
        \midrule
        CPU time [core $\!\cdot\!$ s] & 4415                    & 16260                   & 28605                    \\
        Wall time [s]                 & 1363                    & 4223                    & 7321                     \\
        \bottomrule
    \end{tabularx}
\end{table}

Scaling from 1$\times$ to 5$\times$ raises CPU time by $3.68\times$ and wall time by $3.10\times$; at 10$\times$ the factors are $6.48\times$ and $5.37\times$. Both grow more slowly than the data itself, which is what fixed pipeline and model-setup costs predict, with the compute-intensive training stages accounting for most of the additional cost.

Because each size was measured once and the larger datasets were created by repeating the raw measurement records with shifted timestamps, this is a controlled scaling test rather than a general characterization of larger or more diverse datasets.

\noindent\fbox{%
    \parbox{\dimexpr\linewidth-2\fboxsep-2\fboxrule}{%
        \textit{Answer to RQ4}: Aggregate CPU and wall time grew sublinearly with raw data volume over the tested range, driven by Grid-search and AutoML, while maximum memory rose from 1489~MB to 4027~MB, concentrated in Prepare and Grid-search.
    }%
}

\section{Threats to Validity}

\textit{Internal validity} concerns whether the measured differences result from the frameworks or from their particular implementations. The three pipelines used the same datasets and principal experimental settings, but retained the normal structure and artifact behavior of each implementation. The results therefore describe the complete pipelines and do not isolate orchestration overhead. In particular, LOCALIZE retains additional models and artifacts during Grid-search, while Jupyter and Kedro use the standard \texttt{GridSearchCV} behavior. LOCALIZE measurements also include stage-process startup, whereas Jupyter kernel startup and Kedro initialization before the first catalog load are excluded. This mainly affects the short stages. We implemented all three pipelines, so implementation choices may still favor one approach despite the parity checks. Runtime and memory measurements used the same hardware and matched dependency versions, with five repeated runs, although transient system load may still affect them. The scaling experiment used one run per dataset size.

\textit{Construct validity} concerns whether the selected measures capture the intended properties. LOC favors an implementation when the required capability is already provided, as with AutoML in LOCALIZE, and does not include the original cost of implementing that capability. The four changes cover selected experimental tasks rather than every possible modification. The qualitative grades are based on documentation and hands-on use and remain subject to interpretation. The comparison set is representative rather than exhaustive, and omitted workflow or MLOps platforms may provide overlapping capabilities. Resource usage was sampled at 10\,Hz, so brief fluctuations may not have been captured. Because memory samples from all five runs were combined before calculating the quartiles, the memory error bars describe variation during execution rather than variation between repeated runs.

\textit{External validity} concerns whether the results generalize beyond the studied setting. The framework design is intended to support application-specific ML experimentation beyond localization, while the quantitative evidence in this paper comes from LOCALIZE's radio-localization workflows using tabular wireless-measurement data. The runtime experiments used one CPU-only computing environment and do not characterize other hardware or accelerated workloads. The scaling experiment increased data volume by repeating the raw measurement records with shifted timestamps and used one run per size, so it is a controlled stress test rather than a general characterization of larger or more diverse datasets.

\textit{Reproducibility validity} concerns whether the reported evaluation can be repeated independently. The experiments used versioned code, configurations, data, environment specifications, and fixed random seeds, and the generated Grid-search and AutoML metric summaries were checked for exact consistency across repeated runs within the same controlled environment (repeatability, in the sense of Section~\ref{subsec:methodology-repro}). Numerical values may still differ across hardware, operating systems, or low-level numerical libraries, particularly for AutoML. Custom stages may also introduce new sources of nondeterminism. The implementation, configurations, datasets, environment specifications, and evaluation scripts are publicly available\footnote{\url{https://github.com/sensorlab/LCL-evaluation}} for inspection and replication.

\textit{Design and scope limitations} arise from choices that prioritize simple, reproducible research workflows. LOCALIZE primarily targets tabular radio-localization data and classical machine-learning workflows, with neural models provided through AutoKeras. Supporting other data modalities, learning backends, or optimization frameworks would require new components and may require changes to the implementation.

\section{Conclusion}
\label{sec:conclusion}
This paper addressed how to make repeatable execution a default property of ML experimentation while retaining low-code operation and stage modularity. We proposed a configuration-first design in which experiments are declared in human-readable files, executed as isolated stages with explicit inputs and outputs, and recorded as versioned artifacts. We instantiated the design as LOCALIZE for radio-localization research.

For the targeted workflows, none of the five comparison platforms matched the same combination of low-code operation, common evaluation, stage modularity, and fast iteration (RQ1). LOCALIZE required fewer codebase edits for the dataset, splitting and metric, and AutoML changes, while the estimator change was comparable across implementations (RQ2). Total wall time and peak memory remained comparable despite LOCALIZE's higher Grid-search resource use and process-startup cost in the short stages (RQ3). Over the tested range, resource use grew sublinearly with data volume, consistent with fixed setup and orchestration overhead (RQ4).

The transferable result is the mechanism rather than the tool. Declaring the pipeline, isolating stages behind explicit interfaces, and versioning artifacts are together what produce the recording and cache-aware re-execution behavior, and instantiating the design for another application area then reduces to supplying that area's datasets, data-processing stages, and evaluation procedures. 

Future work should therefore test reproducibility across machines, operating systems, and library versions rather than within a single environment, and instantiate the design in a second application domain.

\section*{Data Availability}
The UMU dataset is available from Zenodo at \url{https://doi.org/10.5281/zenodo.15516876}, and the LOG-a-TEC BLE Fingerprints dataset is available at \url{https://log-a-tec.eu/datasets-ble.html}.

\section*{Code Availability}
The LOCALIZE implementation is available at \url{https://github.com/sensorlab/localize}, and the evaluation code is available at \url{https://github.com/sensorlab/LCL-evaluation}.

\section*{Author Contributions}
Tim Strnad: Software, Writing -- original draft.
Bla\v{z} Bertalani\v{c}: Project administration, Writing -- review \& editing, Validation.
Carolina Fortuna: Funding acquisition, Conceptualization, Writing -- review \& editing.

\section*{Declaration of Generative AI Use}
The authors used ChatGPT for language editing, for identifying candidate references, and for suggesting alternative section outlines. All resulting text was reviewed and revised by the authors, who take full responsibility for the content. The study design, implementation, experiments, and results are the authors' own work and did not involve generative AI.

\section*{Declaration of Competing Interest}
The authors have no relevant financial or non-financial interests to disclose.

\bibliographystyle{IEEEtran}
\bibliography{references}

@misc{huang2018location,
  author    = {Haosheng Huang and Song Gao},
  title     = {Location-Based Services},
  booktitle = {The Geographic Information Science \& Technology Body of Knowledge},
  editor    = {John P. Wilson},
  year      = {2018},
  month     = {March},
  volume    = {2018(Q1)},
  publisher = {The Geographic Information Science \& Technology Body of Knowledge},
  doi       = {10.22224/gistbok/2018.1.14},
  url       = {https://doi.org/10.22224/gistbok/2018.1.14}
}

@article{Pineau2021,
  title     = {Improving Reproducibility in Machine Learning Research},
  author    = {Pineau, Joelle and Vincent-Lamarre, Philippe and Sinha, Koustuv and Lariviere, Vincent and Beygelzimer, Alina and d'Alche-Buc, Florence and Fox, Emily and Larochelle, Hugo},
  journal   = {J. Mach. Learn. Res.},
  volume    = {22},
  number    = {164},
  pages     = {1--20},
  year      = {2021},
  publisher = {JMLR. org}
}

@article{Heil2021,
  title     = {Reproducibility standards for machine learning in the life sciences},
  author    = {Heil, Benjamin J and Hoffman, Michael M and Markowetz, Florian and Lee, Su-In and Greene, Casey S and Hicks, Stephanie C},
  journal   = {Nat. Methods},
  volume    = {18},
  number    = {10},
  pages     = {1132--1135},
  year      = {2021},
  doi       = {10.1038/s41592-021-01256-7},
  publisher = {Nature Publishing Group}
}

@article{Tong2021,
  title     = {{CSI} Fingerprinting Localization With Low Human Efforts},
  author    = {Tong, Xinyu and Wan, Yang and Li, Qianru and Tian, Xiaohua and Wang, Xinbing},
  journal   = {IEEE/ACM Trans. Netw.},
  volume    = {29},
  number    = {1},
  pages     = {372--385},
  year      = {2021},
  doi       = {10.1109/TNET.2020.3035210},
  publisher = {IEEE}
}

@article{gundersen2021do,
  title     = {Do machine learning platforms provide out-of-the-box reproducibility?},
  author    = {Gundersen, Odd Erik and Shamsaliei, Saeid and Isdahl, Richard Juul},
  journal   = {Future Gener. Comput. Syst.},
  volume    = {126},
  pages     = {34--47},
  year      = {2022},
  doi       = {10.1016/j.future.2021.06.014},
  publisher = {Elsevier}
}

@misc{kedro,
  author       = {Alam, Sajid and others},
  title        = {Kedro},
  year         = {2026},
  howpublished = {\url{https://github.com/kedro-org/kedro}},
  note         = {Version 1.5.0}
}

@misc{dvc,
  author    = {{The DVC Team and Contributors}},
  title     = {{DVC}: {Data Version Control} -- {Git} for {Data} and {Models}},
  year      = {2026},
  publisher = {Zenodo},
  note      = {Version 3.67.1},
  doi       = {10.5281/zenodo.19344026},
  url       = {https://doi.org/10.5281/zenodo.19344026}
}

@misc{airflow,
  author       = {{The Apache Software Foundation}},
  title        = {Apache Airflow Documentation},
  year         = {2026},
  howpublished = {\url{https://airflow.apache.org/docs/apache-airflow/stable/}}
}

@misc{prefect,
  author       = {{Prefect Technologies, Inc.}},
  title        = {Prefect Documentation},
  year         = {2026},
  howpublished = {\url{https://docs.prefect.io/v3/get-started}}
}

@misc{luigi,
  author       = {{Spotify AB}},
  title        = {Luigi Documentation},
  year         = {2026},
  howpublished = {\url{https://luigi.readthedocs.io/en/stable/}}
}

@article{snakemake,
  author  = {K{\"o}ster, Johannes and Rahmann, Sven},
  title   = {Snakemake--a Scalable Bioinformatics Workflow Engine},
  journal = {Bioinformatics},
  volume  = {28},
  number  = {19},
  pages   = {2520--2522},
  year    = {2012},
  doi     = {10.1093/bioinformatics/bts480}
}

@article{kreuzberger2023mlops,
  author  = {Kreuzberger, Dominik and K{\"u}hl, Niklas and Hirschl, Sebastian},
  title   = {Machine Learning Operations ({MLOps}): Overview, Definition, and Architecture},
  journal = {IEEE Access},
  volume  = {11},
  pages   = {31866--31879},
  year    = {2023},
  doi     = {10.1109/ACCESS.2023.3262138}
}

@article{zaharia2018mlflow,
  author  = {Zaharia, Matei and Chen, Andrew and Davidson, Aaron and Ghodsi, Ali and Hong, Sue Ann and Konwinski, Andy and Murching, Siddharth and Nykodym, Tomas and Ogilvie, Paul and Parkhe, Mani and Xie, Fen and Zumar, Corey},
  title   = {Accelerating the Machine Learning Lifecycle with {MLflow}},
  journal = {IEEE Data Eng. Bull.},
  volume  = {41},
  number  = {4},
  pages   = {39--45},
  year    = {2018},
  url     = {https://people.eecs.berkeley.edu/~matei/papers/2018/ieee_mlflow.pdf}
}

@article{tagliabue2023metaflow,
  author  = {Tagliabue, Jacopo and Bowne-Anderson, Hugo and Tuulos, Ville and Goyal, Savin and Cledat, Romain and Berg, David},
  title   = {Reasonable Scale Machine Learning with Open-Source {Metaflow}},
  journal = {arXiv preprint arXiv:2303.11761},
  year    = {2023},
  doi     = {10.48550/arXiv.2303.11761}
}

@misc{kubeflow_pipelines,
  author       = {{The Kubeflow Authors}},
  title        = {Kubeflow Pipelines Overview},
  year         = {2026},
  howpublished = {\url{https://www.kubeflow.org/docs/components/pipelines/overview/}}
}

@misc{zenml_stacks,
  author       = {{ZenML GmbH}},
  title        = {{ZenML} Stacks Overview},
  year         = {2026},
  howpublished = {\url{https://docs.zenml.io/stacks}}
}

@article{ye2024mvts,
  title     = {{MvTS}-library: An open library for deep multivariate time series forecasting},
  author    = {Ye, Junchen and Li, Weimiao and Zhang, Zhixin and Zhu, Tongyu and Sun, Leilei and Du, Bowen},
  journal   = {Knowl.-Based Syst.},
  volume    = {283},
  pages     = {111170},
  year      = {2024},
  doi       = {10.1016/j.knosys.2023.111170},
  publisher = {Elsevier}
}

@inproceedings{NETIoT,
  title        = {Towards a Seamless Integration of IoT Devices with IoT Platforms Using a Low-Code Approach},
  author       = {Pantelimon, Silviu-George and Rogojanu, Tudor and Braileanu, Andreea and Stanciu, Valeriu-Daniel and Dobre, Ciprian},
  booktitle    = {2019 IEEE 5th World Forum on Internet of Things (WF-IoT)},
  pages        = {566--571},
  year         = {2019},
  organization = {IEEE},
  doi          = {10.1109/WF-IoT.2019.8767313}
}

@inproceedings{DriotData,
  title        = {ML-Quadrat \& DriotData: A Model-Driven Engineering Tool and a Low-Code Platform for Smart IoT Services},
  author       = {Moin, Armin and Mituca, Andrei and Challenger, Moharram and Badii, Atta and G{\"u}nnemann, Stephan},
  booktitle    = {Proceedings of the 44th International Conference on Software Engineering: Companion Proceedings},
  pages        = {144--148},
  year         = {2022},
  organization = {ACM},
  doi          = {10.1145/3510454.3516841}
}

@inproceedings{arbel_mlxp_2024,
  address    = {New York, NY, USA},
  series     = {{ACM} {REP} '24},
  title      = {{MLXP}: {A} framework for conducting replicable experiments in {Python}},
  isbn       = {979-8-4007-0530-4},
  shorttitle = {{MLXP}},
  url        = {https://dl.acm.org/doi/10.1145/3641525.3663648},
  doi        = {10.1145/3641525.3663648},
  booktitle  = {Proceedings of the 2nd {ACM} {Conference} on {Reproducibility} and {Replicability}},
  publisher  = {Association for Computing Machinery},
  author     = {Arbel, Michael and Zouaoui, Alexander},
  month      = jul,
  year       = {2024},
  pages      = {134--144}
}

@inproceedings{bouckaert_estimating_2004,
  address   = {Banff, Alberta, Canada},
  title     = {Estimating replicability of classifier learning experiments},
  url       = {http://portal.acm.org/citation.cfm?doid=1015330.1015338},
  doi       = {10.1145/1015330.1015338},
  language  = {en},
  urldate   = {2025-02-13},
  booktitle = {Twenty-first international conference on {Machine} learning  - {ICML} '04},
  publisher = {ACM Press},
  author    = {Bouckaert, Remco R.},
  year      = {2004},
  pages     = {15}
}

@article{gonzalezcebrian_standardised_2024,
  title      = {Standardised {Versioning} of {Datasets}: a {FAIR}–compliant {Proposal}},
  volume     = {11},
  issn       = {2052-4463},
  shorttitle = {Standardised {Versioning} of {Datasets}},
  url        = {https://www.nature.com/articles/s41597-024-03153-y},
  doi        = {10.1038/s41597-024-03153-y},
  language   = {en},
  number     = {1},
  urldate    = {2025-02-13},
  journal    = {Sci. Data},
  author     = {González–Cebrián, Alba and Bradford, Michael and Chis, Adriana E. and González–Vélez, Horacio},
  month      = apr,
  year       = {2024},
  pages      = {358}
}

@article{zhuang_easyfl_2022,
  title      = {{EasyFL}: {A} {Low}-{Code} {Federated} {Learning} {Platform} for {Dummies}},
  volume     = {9},
  issn       = {2327-4662},
  shorttitle = {{EasyFL}},
  url        = {https://ieeexplore.ieee.org/document/9684558/?arnumber=9684558},
  doi        = {10.1109/JIOT.2022.3143842},
  number     = {15},
  urldate    = {2025-02-13},
  journal    = {IEEE Internet Things J.},
  author     = {Zhuang, Weiming and Gan, Xin and Wen, Yonggang and Zhang, Shuai},
  month      = aug,
  year       = {2022},
  pages      = {13740--13754}
}

@inproceedings{isdahl_out---box_2019,
  address    = {San Diego, CA, USA},
  title      = {Out-of-the-{Box} {Reproducibility}: {A} {Survey} of {Machine} {Learning} {Platforms}},
  copyright  = {https://ieeexplore.ieee.org/Xplorehelp/downloads/license-information/IEEE.html},
  isbn       = {978-1-7281-2451-3},
  shorttitle = {Out-of-the-{Box} {Reproducibility}},
  url        = {https://ieeexplore.ieee.org/document/9041744/},
  doi        = {10.1109/eScience.2019.00017},
  language   = {en},
  urldate    = {2025-02-14},
  booktitle  = {2019 15th {International} {Conference} on {eScience} ({eScience})},
  publisher  = {IEEE},
  author     = {Isdahl, Richard and Gundersen, Odd Erik},
  month      = sep,
  year       = {2019},
  pages      = {86--95}
}

@article{trevlakis_localization_2023,
  title      = {Localization as a {Key} {Enabler} of {6G} {Wireless} {Systems}: {A} {Comprehensive} {Survey} and an {Outlook}},
  issn       = {2644-125X},
  shorttitle = {Localization as a {Key} {Enabler} of {6G} {Wireless} {Systems}},
  url        = {https://ieeexplore.ieee.org/ielx7/8782661/10008219/10287134.pdf},
  doi        = {10.1109/OJCOMS.2023.3324952},
  journal    = {IEEE Open J. Commun. Soc.},
  author     = {Trevlakis, Stylianos E. and Boulogeorgos, Alexandros-Apostolos A. and Pliatsios, Dimitrios and Querol, Jorge and Ntontin, Konstantinos and Sarigiannidis, Panagiotis and Chatzinotas, Symeon and Di Renzo, Marco},
  volume     = {4},
  pages      = {2733--2801},
  year       = {2023}
}

@article{semmelrock_reproducibility_2025,
  author   = {Semmelrock, Harald and Ross-Hellauer, Tony and Kopeinik, Simone and Theiler, Dieter and Haberl, Armin and Thalmann, Stefan and Kowald, Dominik},
  title    = {Reproducibility in machine-learning-based research: Overview, barriers, and drivers},
  journal  = {AI Mag.},
  volume   = {46},
  number   = {2},
  pages    = {e70002},
  doi      = {10.1002/aaai.70002},
  url      = {https://onlinelibrary.wiley.com/doi/abs/10.1002/aaai.70002},
  year     = {2025}
}

@article{henderson_deep_2018,
  title   = {Deep reinforcement learning that matters},
  author  = {Henderson, Peter and Islam, Riashat and Bachman, Philip and Pineau, Joelle and Precup, Doina and Meger, David},
  journal = {Proc. AAAI Conf. Artif. Intell.},
  volume  = {32},
  number  = {1},
  year    = {2018},
  pages   = {3207--3214},
  doi     = {10.1609/aaai.v32i1.11694}
}

@inproceedings{bouthillier_accounting_2021,
  author    = {Bouthillier, Xavier and Delaunay, Pierre and Bronzi, Mirko and Trofimov, Assya and Nichyporuk, Brennan and Szeto, Justin and Mohammadi Sepahvand, Nazanin and Raff, Edward and Madan, Kanika and Voleti, Vikram and Ebrahimi Kahou, Samira and Michalski, Vincent and Arbel, Tal and Pal, Chris and Varoquaux, Gael and Vincent, Pascal},
  booktitle = {Proceedings of Machine Learning and Systems},
  editor    = {A. Smola and A. Dimakis and I. Stoica},
  pages     = {747--769},
  title     = {Accounting for Variance in Machine Learning Benchmarks},
  url       = {https://proceedings.mlsys.org/paper_files/paper/2021/file/0184b0cd3cfb185989f858a1d9f5c1eb-Paper.pdf},
  volume    = {3},
  year      = {2021}
}

@inproceedings{pham_problems_nodate,
  author    = {Pham, Hung Viet and Qian, Shangshu and Wang, Jiannan and Lutellier, Thibaud and Rosenthal, Jonathan and Tan, Lin and Yu, Yaoliang and Nagappan, Nachiappan},
  title     = {Problems and opportunities in training deep learning software systems: an analysis of variance},
  year      = {2021},
  isbn      = {9781450367684},
  publisher = {Association for Computing Machinery},
  address   = {New York, NY, USA},
  url       = {https://doi.org/10.1145/3324884.3416545},
  doi       = {10.1145/3324884.3416545},
  booktitle = {Proceedings of the 35th IEEE/ACM International Conference on Automated Software Engineering},
  pages     = {771–783},
  numpages  = {13},
  location  = {Virtual Event, Australia},
  series    = {ASE '20}
}

@misc{frank_weka_workbench_2016,
  author       = {Frank, Eibe and Hall, Mark A. and Witten, Ian H.},
  title        = {The {WEKA} Workbench},
  year         = {2016},
  howpublished = {Online appendix for \emph{Data Mining: Practical Machine Learning Tools and Techniques}, 4th ed., Morgan Kaufmann},
  publisher    = {University of Waikato},
  url          = {https://hdl.handle.net/10289/17096}
}

@article{cardoso_monai_2022,
  author  = {Cardoso, M. Jorge and Li, Wenqi and Brown, Richard and others},
  title   = {{MONAI}: An Open-Source Framework for Deep Learning in Healthcare},
  journal = {arXiv preprint arXiv:2211.02701},
  year    = {2022},
  doi     = {10.48550/arXiv.2211.02701},
  url     = {https://arxiv.org/abs/2211.02701}
}

@article{orange_pdf,
  title    = {Orange: {Data} {Mining} {Toolbox} in {Python}},
  language = {en},
  journal  = {J. Mach. Learn. Res.},
  volume   = {14},
  number   = {71},
  pages    = {2349--2353},
  year     = {2013},
  author   = {Dem{\v{s}}ar, Janez and Curk, Toma{\v{z}} and Erjavec, Ale{\v{s}} and others}
}

@article{granger_jupyter_2021,
  title      = {Jupyter: {Thinking} and {Storytelling} {With} {Code} and {Data}},
  volume     = {23},
  issn       = {1558-366X},
  shorttitle = {Jupyter},
  url        = {https://ieeexplore.ieee.org/document/9387490},
  doi        = {10.1109/MCSE.2021.3059263},
  number     = {2},
  urldate    = {2025-09-16},
  journal    = {Comput. Sci. Eng.},
  author     = {Granger, Brian E. and Pérez, Fernando},
  month      = mar,
  year       = {2021},
  pages      = {7--14}
}

@misc{ludwig_pdf,
  author = {Piero Molino and Yaroslav Dudin and Sai Sumanth Miryala},
  title  = {Ludwig: a type-based declarative deep learning toolbox},
  year   = {2019},
  eprint = {1909.07930},
  doi    = {10.48550/arXiv.1909.07930},
  url    = {https://arxiv.org/abs/1909.07930}
}

@article{matlab_survey,
  title    = {Indoor positioning with Wi-Fi Location: A survey of IEEE 802.11mc/az/bk fine timing measurement research},
  journal  = {Comput. Commun.},
  volume   = {247},
  pages    = {108400},
  year     = {2026},
  issn     = {0140-3664},
  doi      = {10.1016/j.comcom.2025.108400},
  url      = {https://www.sciencedirect.com/science/article/pii/S0140366425003573},
  author   = {Katarzyna Kosek-Szott and Szymon Szott and Wojciech Ciezobka and Maksymilian Wojnar and Krzysztof Rusek and Jonathan Segev}
}

@article{Roy2021MLSurvey,
  title   = {A Survey of Machine Learning Techniques for Indoor Localization and Navigation Systems},
  author  = {Priya Roy and Chandreyee Chowdhury},
  journal = {J. Intell. Robot. Syst.},
  volume  = {101},
  number  = {3},
  pages   = {63},
  year    = {2021},
  doi     = {10.1007/s10846-021-01327-z}
}

@article{python_survey,
  author  = {Sebastian Raschka and Joshua Patterson and Corey Nolet},
  title   = {Machine Learning in Python: Main Developments and Technology Trends in Data Science, Machine Learning, and Artificial Intelligence},
  journal = {Information},
  volume  = {11},
  number  = {4},
  pages   = {193},
  year    = {2020},
  doi     = {10.3390/info11040193}
}

@inproceedings{morano_outdoor_2022,
  title        = {LOG-a-TEC Testbed outdoor localization using BLE beacons},
  author       = {Bertalani{\v{c}}, Bla{\v{z}} and Morano, Grega and Cerar, Gregor},
  booktitle    = {2022 International Balkan Conference on Communications and Networking (BalkanCom)},
  pages        = {115--119},
  year         = {2022},
  organization = {IEEE},
  doi          = {10.1109/BalkanCom55633.2022.9900607}
}

@misc{asensio-garriga_university_2025,
  title     = {University of {Murcia} {5G} {Dataset} 1},
  url       = {https://zenodo.org/records/15516876},
  urldate   = {2025-09-15},
  publisher = {Zenodo},
  author    = {Asensio-Garriga, Rodrigo and Pogo Medina, Anthony Joel and Alarcon-Hellin, Gonzalo and Bernal Escobedo, Luis and Sanchez-Iborra, Ramon and Skarmeta Gómez, Antonio},
  month     = may,
  year      = {2025},
  doi       = {10.5281/zenodo.15516876}
}

@inproceedings{gauger_massive_2020,
  title     = {Massive {MIMO} {Channel} {Measurements} and {Achievable} {Rates} in a {Residential} {Area}},
  url       = {https://ieeexplore.ieee.org/abstract/document/9097159},
  urldate   = {2025-09-16},
  booktitle = {{WSA} 2020; 24th {International} {ITG} {Workshop} on {Smart} {Antennas}},
  author    = {Gauger, Marc and Arnold, Maximilian and ten Brink, Stephan},
  month     = feb,
  publisher = {IEEE},
  address   = {Hamburg, Germany},
  year      = {2020},
  pages     = {1--6}
}

@inproceedings{arnold_novel_2019,
  title     = {Novel {Massive} {MIMO} {Channel} {Sounding} {Data} applied to {Deep} {Learning}-based {Indoor} {Positioning}},
  url       = {https://ieeexplore.ieee.org/abstract/document/8661318},
  doi       = {10.30420/454862021},
  urldate   = {2025-09-16},
  booktitle = {{SCC} 2019; 12th {International} {ITG} {Conference} on {Systems}, {Communications} and {Coding}},
  author    = {Arnold, Maximilian and Hoydis, Jakob and Brink, Stephan ten},
  month     = feb,
  year      = {2019},
  publisher = {IEEE},
  address   = {Hamburg, Germany},
  pages     = {1--6}
}

@inproceedings{narayanan_lumos5g_2020,
  address    = {Virtual Event USA},
  title      = {{Lumos5G}: {Mapping} and {Predicting} {Commercial} {mmWave} {5G} {Throughput}},
  isbn       = {978-1-4503-8138-3},
  shorttitle = {{Lumos5G}},
  url        = {https://dl.acm.org/doi/10.1145/3419394.3423629},
  doi        = {10.1145/3419394.3423629},
  language   = {en},
  urldate    = {2025-09-16},
  booktitle  = {Proceedings of the {ACM} {Internet} {Measurement} {Conference}},
  publisher  = {ACM},
  author     = {Narayanan, Arvind and Ramadan, Eman and Mehta, Rishabh and Hu, Xinyue and Liu, Qingxu and Fezeu, Rostand A. K. and Dayalan, Udhaya Kumar and Verma, Saurabh and Ji, Peiqi and Li, Tao and Qian, Feng and Zhang, Zhi-Li},
  month      = oct,
  year       = {2020},
  pages      = {176--193}
}

@article{wilson2017good,
  title     = {Good enough practices in scientific computing},
  author    = {Wilson, Greg and Bryan, Jennifer and Cranston, Karen and Kitzes, Justin and Nederbragt, Lex and Teal, Tracy K},
  journal   = {PLoS Comput. Biol.},
  volume    = {13},
  number    = {6},
  pages     = {e1005510},
  year      = {2017},
  publisher = {Public Library of Science San Francisco, CA USA},
  doi       = {10.1371/journal.pcbi.1005510}
}

@article{dolata2023making,
  title     = {Making sense of {AI} systems development},
  author    = {Dolata, Mateusz and Crowston, Kevin},
  journal   = {IEEE Trans. Softw. Eng.},
  volume    = {50},
  number    = {1},
  pages     = {123--140},
  year      = {2024},
  doi       = {10.1109/TSE.2023.3338857},
  publisher = {IEEE}
}

@article{10855627,
  author   = {Zhao, Zhimin and Bangash, Abdul Ali and Côgo, Filipe Roseiro and Adams, Bram and Hassan, Ahmed E.},
  journal  = {IEEE Trans. Softw. Eng.},
  title    = {On the Workflows and Smells of Leaderboard Operations (LBOps): An Exploratory Study of Foundation Model Leaderboards},
  year     = {2025},
  volume   = {51},
  number   = {4},
  pages    = {929--946},
  doi      = {10.1109/TSE.2025.3533972}
}

@article{makke2024interpretable,
  title     = {Interpretable scientific discovery with symbolic regression: a review},
  author    = {Makke, Nour and Chawla, Sanjay},
  journal   = {Artif. Intell. Rev.},
  volume    = {57},
  number    = {1},
  pages     = {2},
  year      = {2024},
  publisher = {Springer},
  doi       = {10.1007/s10462-023-10622-0}
}

@misc{acm_badging_2020,
  title        = {Artifact Review and Badging -- Version 1.1},
  author       = {{Association for Computing Machinery}},
  year         = {2020},
  howpublished = {\url{https://www.acm.org/publications/policies/artifact-review-and-badging-current}},
  note         = {Terminology aligned with the U.S. National Academies (2019)}
}

@article{10.1145/240455.240464,
author = {Fayyad, Usama and Piatetsky-Shapiro, Gregory and Smyth, Padhraic},
title = {The KDD process for extracting useful knowledge from volumes of data},
year = {1996},
issue_date = {Nov. 1996},
publisher = {Association for Computing Machinery},
address = {New York, NY, USA},
volume = {39},
number = {11},
issn = {0001-0782},
url = {https://doi.org/10.1145/240455.240464},
doi = {10.1145/240455.240464},
journal = {Commun. ACM},
month = nov,
pages = {27–34},
numpages = {8}
}

\begin{IEEEbiography}[{\includegraphics[width=1in,height=1.25in,clip,keepaspectratio]{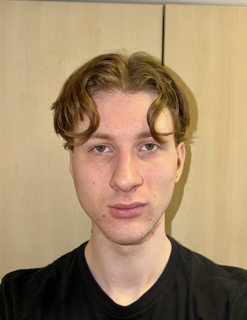}}]{TIM STRNAD}
is a student in the interdisciplinary Computer Science and Mathematics programme at the Faculty of Computer and Information Science, University of Ljubljana. He currently works at SensorLab, Jo\v{z}ef Stefan Institute, as a summer student, contributing to research software for reproducible machine-learning experimentation and radio localization.
\end{IEEEbiography}

\begin{IEEEbiography}[{\includegraphics[width=1in,height=1.25in,clip,keepaspectratio]{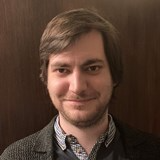}}]{BLA\v{Z} BERTALANI\v{C}}
received his Ph.D. degree with highest distinction from the Faculty of Electrical Engineering, University of Ljubljana. He is a researcher at SensorLab, Jo\v{z}ef Stefan Institute. His research interests include machine learning and artificial intelligence, particularly time-series analysis and smart infrastructures. He is an IEEE member, has held several leadership positions in the IEEE Slovenia Section, and has authored more than 15 IEEE publications.
\end{IEEEbiography}

\begin{IEEEbiography}[{\includegraphics[width=1in,height=1.25in,clip,keepaspectratio]{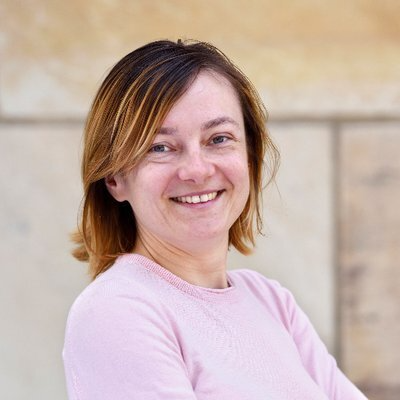}}]{CAROLINA FORTUNA}
has more than 15 years of experience at the intersection of artificial intelligence and communication networks. She is a Research Associate Professor at the Jo\v{z}ef Stefan Institute, where she leads SensorLab. She previously conducted research at Ghent University and Stanford University and has led teams in numerous EU-funded projects. She has advised or co-advised more than 10 M.Sc. and Ph.D. students and coauthored more than 100 peer-reviewed papers. Her research focuses on next-generation smart infrastructures that improve everyday life.
\end{IEEEbiography}

\EOD

\end{document}